# Highly Efficient and Selective Extraction of Gold by Reduced Graphene Oxide

Fei Li[#,1], Jiuyi Zhu[#,1], Pengzhan Sun[2], Mingrui Zhang[1], Zhenqing Li[1], Dingxin Xu[1], Xinyu Gong[1], Xiaolong Zou[1], A. K. Geim[1,2,*], Yang Su[1,*], Hui-Ming Cheng[3,4,*]

[1]Tsinghua-Berkeley Shenzhen Institute & Institute of Materials Research, Tsinghua Shenzhen International Graduate School, Tsinghua University, Shenzhen 518055, P. R. China

[2]School of Physics & Astronomy, University of Manchester, Manchester M13 9PL, United Kingdom

[3]Faculty of Materials Science and Engineering / Institute of Technology for Carbon Neutrality, Shenzhen Institute of Advanced Technology, Chinese Academy of Sciences, Shenzhen 518055, P. R. China

[4]Shenyang National Laboratory for Materials Sciences, Institute of Metal Research, Chinese Academy of Sciences, Shenyang 110016, P. R. China

[#]These authors contributed equally to this work.

*Corresponding authors: Andre.K.Geim@manchester.ac.uk (A.K.Geim); Su.yang@sz.tsinghua.edu.cn (Y. Su); Cheng@imr.ac.cn (H.-M. Cheng)

**Abstract**

Materials that are capable of extracting gold from complex sources, especially electronic waste (e-waste) with high efficiency are needed for gold resource sustainability and effective e-waste recycling. However, it remains challenging to achieve high extraction capacity to trace amount of gold, and precise selectivity to gold over a wide range of complex co-existing elements. Here we report a reduced graphene oxide (rGO) material that has an ultrahigh extraction capacity for trace amounts of gold (1,850 mg/g and 1,180 mg/g to 10 ppm and 1 ppm gold). The excellent gold extraction behavior is accounted to the graphene areas and oxidized regions of rGO. The graphene areas spontaneously reduce gold ions to metallic gold, and the oxidized regions provide a good dispersibility so that efficient adsorption and reduction of gold ions by the graphene area can be realized. The rGO is also highly selective to gold ions. By controlling the protonation process of the functional groups on the oxidized regions of rGO, it shows an exclusive gold extraction without adsorption of 14 co-existing elements seen in e-waste. These discoveries are further exploited in highly efficient, continuous gold recycling from e-waste with good scalability and economic viability, as exemplified by extracting gold from e-waste using a rGO membrane based flow-through process.

**Introduction**

Electronic waste (e-waste) is the world's fastest-growing solid waste and poses risks to the environment and human health. Less than 20% of e-waste has currently been recycled, primarily for a lack of technologies with sufficient efficiency and economic viability to recover valuable elements within it[1-5]. Gold is the most valuable part in e-waste, and its efficient extraction can turn this recycling challenge into profitable business[6-8]. Activated carbon is widely used for gold extraction but has significant drawbacks including low extraction capacity, poor selectivity, and high energy and resource intensity[9-11]. There is a strong demand to develop novel gold extraction materials with higher extraction capacity and selectivity. Various novel gold adsorbents have been explored recently. These gold adsorbents could be categorized into two sets, one is nanoporous materials, for example, metal-organic framework[12,13], covalent organic polymer[4,14], and porous aromatic framework[15]. The gold extraction behavior of these materials is mainly contributed by the immobilization of gold ions with the intrinsic porosity and the chemical reduction of the gold ion by the added functional groups. The other set of gold adsorbents, for example, two-dimensional molybdenum disulphide[16-18], amyloid[19,20], cyclodextrin[21] and diamide[22], instead of relying on the porosity of the adsorbents, their efficient gold extraction is accounted to the chemical reduction of gold ions to $Au^0$ by photoreduction or the functional groups, and precipitation of gold ions with the adsorbents.

These gold adsorbents exhibit a high gold extraction capacity at a gold concentration from 500 ppm-3000 ppm[12-22], but this capacity decreases to less than 250 mg/g at a gold concentration range relevant to e-waste recycling, specifically, from ppb level to tens of ppm[4,12,19]. Furthermore, as the e-waste contains complex co-existing metal elements, the practical gold extraction process often requires good extraction selectivity to gold, so that separation of co-existing elements from the extracted gold can be avoided, achieving a high resource- and energy- efficiency for gold recycling. The existing novel gold adsorbents have demonstrated good selectivity to gold, but their adsorption to the co-existing metal elements is still unneglectable[4,12,15,16,19]. Therefore, the development of materials with high gold extraction capacity to trace amount of gold, precise gold selectivity and economic viability remains lacking.

Here we report an exceptionally high gold extraction capacity of chemically reduced graphene oxide (rGO), reaching 1,850 mg/g and 9,059 mg/g when extracting gold from its 10 ppm solution at 25 °C and 60 °C, respectively, combined with an ability to extract gold at minute concentrations, down to parts per trillion and high selectivity. During extraction, rGO reduces >95% gold ions to metallic gold, avoiding elution and precipitation necessary in post-adsorption processing. Moreover, gold extraction can be done selectively, without adsorption of the other 14 elements normally present in e-waste. This in turn enables the recycling of copper, the second valuable metal in e-waste. Finally, we demonstrate a highly efficient flow-through process for gold extraction using rGO membranes. Our findings show a promising venue for addressing global e-waste challenges and gold scarcity.

## Results and Discussion
### Highly efficient gold extraction by rGO

The rGO was obtained by chemical reduction of commercial, mass-produced graphene oxide (GO) nanosheets using ascorbic acid as a reductant (Methods). The resulting rGO nanosheets were repeatedly washed with deionized water, until a stable colloid was formed, and then the rGO suspension was added directly into solutions containing gold ions (Fig. 1a). Considering the hydrochemical process is a flexible, low-cost, and sustainable method for e-waste recycling[23,24], and [AuCl$_4$]$^-$ is a common gold complex seen in a hydrochemical process [4,7,12], we have chosen KAuCl$_4$ solutions prepared in different concentrations ($C$) to examine the gold extraction capacity of rGO (Fig.1a and b); Similar to the behaviour observed for other adsorbents[4,13,14], the extraction capacity increased with increasing $C$ and reached a plateau above 10 ppm (Fig. 1b). Specifically, 1 g of rGO can extract ~ 690, 1,180, and 1,850 mg of gold from 0.1, 1 and 10 ppm gold solutions, respectively. This extraction capacity outperformed those of novel nanoporous adsorbents at same concentrations and, importantly, extends to solutions containing only 1 ppb of gold (Supplementary Note 1; Supplementary Fig. 1 and Supplementary Table 1). Although we could not quantify the extraction capacity for even lower $C$ (our quantitative methods reached their detection limit), we still observed significant gold extraction down to as little as 10 ppt (Supplementary Fig. 1). This makes the reported rGO technique particularly interesting because wastewater and freshwater often contain gold with $C$ below 10 ppb, and its efficient reclaim is challenging[12].

When measuring the extraction capacity of rGO to different pH from 2 to 11, we found the extraction capacity varied with the pH, which is not unusual and similar to many gold adsorbents[4,13-16,19-21]. As shown in Supplementary Fig. 2, the maximum gold uptake was found at pH ≈ 4. This is attributed to the fact that, at lower pH, the rGO colloid loses its stability[25,26] whereas, at basic pH, rGO undergoes deprotonation, becomes negatively charged[4] and starts repelling negatively charged $[AuCl_4]^-$. Both effects decrease the extraction capacity. It is important to note that, the extraction capacities at the acidic and basic solution are still higher than the previously reported gold extraction capacities at the same $C$ (Supplementary Fig. 1 and 2), and by increasing the amount of rGO, ~99% gold ions from its 10 ppm solution with the pH of 2-11 can be extracted (Methods and Fig. 1b). We also investigated the gold extraction as a function of time (Fig. 1c and Supplementary Fig. 3). Within 10 minutes, 1 g of rGO provided extraction of ~1,010 and 325 mg of gold from 10 ppm and 100 ppb solutions, respectively (Fig. 1c), suggesting not only efficient but also rapid extraction. Interestingly, after extraction to 1 ppm gold, rGO changed its colour from black to gold within hours, suggesting almost instantaneous Au recovery (Supplementary Movie 1).

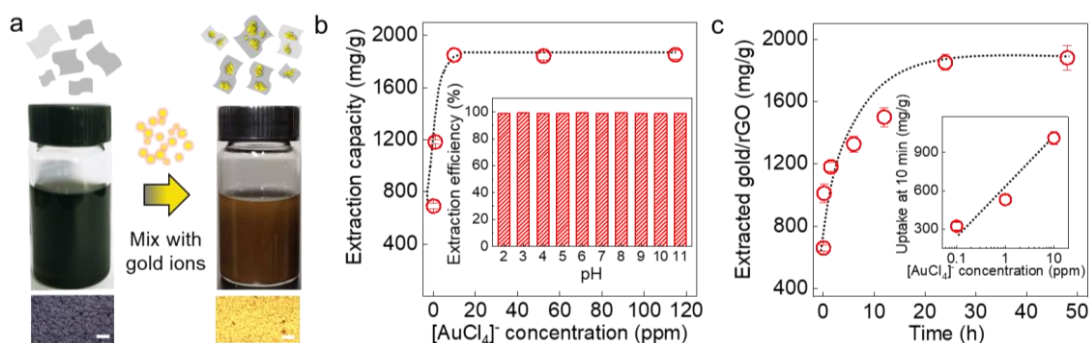

**Fig. 1| High-efficiency extraction of trace amounts of gold.** (**a**) Schematic of the extraction process using rGO. After mixed with gold ion (10 ppm) for 12 h, rGO suspension gradually changed its color from black to brown. The bottom panels are the optical images of rGO films deposited from rGO nanosheets before (black) and after gold extraction (gold). Scale bar: 20 μm. (**b**) Extraction capacity as a function of gold concentration after 24 h. The inset plots the extraction efficiency measured for 10 ppm solutions at different pH. (**c**) Extraction capacity as a function of time for a 10 ppm gold solution. Inset: Extraction from different solutions after 10 minutes. Dashed lines, guides to the eye. All the experiments were performed at 25 ºC. All the error bars in this figure represent the standard deviation.

**Mechanism study of rGO's gold extraction behavior**

To understand the ultrahigh gold extraction behavior of rGO, we started with the characterization of rGO after the gold extraction. We found nanoparticles on top of rGO nanosheets after 24 h of its exposure to gold solution (Fig. 2a). X-ray diffraction (XRD) showed this particulate to be metallic gold, suggesting that $[AuCl_4]^-$ was reduced to $Au^0$ (Supplementary Fig. 4). This was confirmed by X-ray photoelectron spectroscopy (XPS) analysis (Fig. 2b) which showed that >95% of the particulate was metallic $Au^0$ rather than $[AuCl_4]^-$. Thermal analysis (Supplementary Note 2; Supplementary Fig. 4) provided good agreement with the XPS results, as no endothermic peak for the transformation of gold salts into $Au^0$ was found by differential scanning calorimetry (DSC). These observations suggest that a reductive adsorption mechanism dominates the extraction process and, if any adsorption of $[AuCl_4]^-$ on rGO occurs, its contribution is minor. Furthermore, we found that gold nanoparticles formed already after 2 minutes of the contact of rGO with gold solutions (Supplementary Note 3; Supplementary Fig. 7). Moreover, DSC analysis after 10 minutes showed no peak attributable to the gold salt precursor (Supplementary Fig. 8). This suggests that the reductive adsorption of gold ions happens throughout the entire extraction process, including its initial stages. In comparison to the reported novel gold adsorbents, though some showed reductive adsorption, other physi- and chemi-sorptions were also significant. That is, gold extracted by these adsorbents, was a mixture of ionic and metallic gold [4,12-15,19,27-29], therefore, it would require energy- and cost-intensive post-processing, for example, elution and precipitation to desorb and reduce the ionic gold, so that a full adsorption capacity can be achieved[30,31]. In contrast, as >95% extracted Au ions was reduced to metallic gold by rGO, it allowed direct isolation of metallic gold from rGO surface without further post-adsorption processing, providing an extra advantage of rGO for gold extraction.

Because no extra reductant for gold ions was used during gold extraction, it is sensible to conclude that the observed reductive adsorption was a redox reaction between the gold ions and rGO. We have performed gold extraction at different temperatures ($T$) to investigate the kinetics of this redox reaction. As the gold extraction capacity is a direct result of redox reaction rate, we have plotted the extraction capacity versus $T$, and fitted with Arrhenius equation, $\exp(-E/k_BT)$, that $E$ is the energy barrier and $k_B$ is the Boltzmann constant. This yielded an energy barrier 0.12±0.02 eV (Fig. 2c), suggesting an activation behaviour for gold extraction, which is consistent with our DSC analysis that gold reduction by rGO is an endothermic process (Supplementary Fig. 4), the $E$ is much lower than the activation energy seen in typical chemisorption process (0.21 eV-4.3 eV)[32], supporting the observed fast and ultrahigh extraction capacity at low $C$. Furthermore, the observed temperature dependence allowed improving gold extraction capacity by increasing $T$, for example, at $T$= 45 °C and 60 °C, we observed extraction capacities ($C$ =10 ppm) of 4,100 mg/g and 9,059 mg/g respectively.

Next, we studied the influences of the redox pair, gold ion and rGO on the gold extraction performance. For the gold ions, we have measured rGO's adsorption capacity to gold ions with different reduction potential $E_0$ (Fig. 2d). Similar capacities were found for solutions of $Au^{3+}$, $[AuCl_4]^-$ and $[AuBr_4]^-$. Note that these complexes were most frequently used for gold extraction by novel adsorbents as well as for doping of graphene by gold-containing solutions[4,12,14,21,33,34]. This also suggested the possibility for rGO to extract gold from gold solutions containing different gold ions. As a demonstration, we found rGO reclaimed ~100% gold from 10 ppm $[AuBr_4]^-$ solution obtained by dissolving gold in aqueous solution of N-bromosuccinimide at pH=~8[35]. For ions with $E_0$ < 0.8 eV, for example, $[Au(CN)_2]^-$ and $[Au(S_2O_3)_2]^{3-}$, rGO showed little extraction, corroborating our conclusion that the redox reaction between gold ion and rGO dictated the reported behaviour.

To study the influence of rGO on the redox reaction, we recalled the local atomic structure of rGO nanosheet (Fig. 2h), which contains large unoxidized (or reduced) graphene areas, some oxidized regions that provide stability of rGO dispersions, and sub-nanometre pinholes[36]. Given a little fraction of pinhole (~5%) on rGO[36], we only focused on the contribution of graphene area and oxidized regions to the gold extraction behavior.

For the graphene area, in previous reports of gold ion ($Au^{3+}$ or $[AuCl_4]^-$) doped graphene grown by chemical vapour deposition (CVD), the graphene was found to chemically reduce these gold ions via a redox reaction mechanism, in which, graphene donated electrons to reduce $Au^{3+}$ to $Au^0$.[37,38] Though these studies only focused on a high C (≥200 ppm) and a short doping time (one to few minutes), they implied the observed reductive adsorption likely was accounted to the graphene area.

To validate that the graphene area of rGO reduced the gold ion, we have firstly studied the gold reduction by pristine graphene obtained by mechanical exfoliation, so that possible influence of the oxidized region can be excluded. The gold nanoparticles formed on the graphene after a few minutes and continued to grow even after many hours[33,34]. Using areal extraction capacity allowed us to quantify the gold extraction capability of mechanically exfoliated graphene. The monolayer graphene showed the highest areal extraction capacity, in quantitative agreement with the weight capacity observed for rGO (Supplementary Note 4), suggesting the graphene area in rGO was responsible for ultrahigh gold extraction capacity. Rather unexpectedly, as shown in Fig. 2e, increasing the layer number of graphene *(N)*, the areal gold extraction capacity, which should be independent to $N$ as only the outmost graphene layer had a contact with the gold ion and there's no gold intercalation between graphene layer, instead, decreased with $N$, indicating not only the redox potential between graphene and gold ions is essential but also other factors are involved in the process. Further SEM analysis revealed that wrinkles and folds frequently seen for mono- and few-layer graphene tended to accumulate more gold nanoparticles than flat areas (Supplementary Fig. 9 and 10), suggesting a higher chemical activity for gold extraction at the warped graphene, which was in agreement with previous reported gold doped CVD graphene[37], this can be explained by our simulation (Supplementary Fig. 11) that the strain generated in the warped area of graphene, decreased the adsorption energy to gold ion and enhanced electron transfer of graphene to gold ion, both were beneficial for reductive adsorption of gold ion (Supplementary Note 4).

Further insights on the contribution of graphene area to the gold reduction were obtained by considering the electron transfer from the graphene area to gold ions, the Raman and ultraviolet-visible (UV-Vis) spectroscopy were used to probe the electron transfer of rGO by comparing the spectra before and after gold extraction. The Raman spectrum showed that after extraction, the rGO became p-doped and showed an increased number of defects in the rGO nanosheets[39] (Supplementary Fig. 5a and Supplementary Fig. 6). The UV-Vis spectroscopy revealed that, the characteristic peak of rGO shifted towards that of GO (Supplementary Fig. 5b). These two pieces of evidence indicated the donation of electrons from rGO for gold reduction[40], confirming the reduction of gold ions by the graphene area of rGO. In another set of experiments, encouraged by the fact that graphene reductively adsorbed gold ions, the gold extraction capacities of commercial graphene and expanded graphite were measured. Both graphene materials were known to have well-retained graphene areas but little oxygen-containing functional group, which was also verified by the Raman analysis (Supplementary Note 5, Supplementary Fig. 12). During the extraction, they showed no dispersibility in gold solution because of their hydrophobicity, and their extraction capacities were <100 mg/g (Fig. 2g).

The low extraction capacity observed in commercial graphene suggested graphene area was not the only factor contributing to the ultrahigh extraction capacity of rGO. Thus, we studied the influence of the other atomic structure, oxidized regions of rGO, on the extraction performance. Given the synthesis of rGO was a process to remove some oxidized regions on GO (Fig. 2h), we have controlled the reduction time of GO, and XPS analysis confirmed the oxidized regions were gradually removed with the reduction time (Supplementary Note 5, Supplementary Fig. 12a). As shown in Fig. 2g, at a $C=$ 10 ppm, pristine GO showed little gold extraction capacity, but the rGO obtained by 10-minute reduction exhibited an extraction capacity ~1,100 mg/g, and 30-minute reduction gave the highest extraction capacity ~1,850 mg/g (Fig. 2f), which could be explained by the increased graphene area in the rGO with a prolonged reduction time. Therefore, more sites were available for reductive adsorption. Further increasing the reduction time to 1 and 4 h, rGO agglomerated as validated by their decreased zeta potentials, and the capacities have decreased to ~1,460 and 1,180 mg/g, respectively. This can be understood that, the reduction was beneficial for the recovery of graphene areas of rGO, but a long-time reduction excessively eliminated the oxidized regions. As a result, agglomeration and a poor dispersibility of the rGO was seen, such agglomeration of rGO disabled some of the recovered graphene areas from being exposed to the aqueous gold ion, and inhibited the reductive adsorption, leading to a low extraction capacity (Fig. 2h). In addition, we also have tried rGO reduced by hydrazine or hydroquinone for gold extraction. Both rGO had recovered graphene area and formed stable monolayer dispersions as confirmed by their zeta potentials (Supplementary Note 5; Supplementary Fig. 12). They exhibited high extraction capacities (~1,190 and 1,835 mg/g from $C$ = 10 ppm, respectively) as shown in Fig. 2g, similar to the values for the rGO reduced in ascorbic acid. These experiments unambiguously showed that, both the graphene area and the oxidized regions were critical for the observed high gold extraction capacity.

With the above experimental results, we proposed the mechanism for the ultrahigh gold extraction of rGO as follows, firstly, the adsorption was driven by the concentration difference ($\Delta C$) of adsorbates in its solution and on the adsorbent[41]. In our case, $\Delta C = C_{sol}^{Au\ ion} - C_{rGO}^{Au\ ion}$, Once gold ions were adsorbed on rGO, >95% of them were reduced to $Au^0$ spontaneously. Such rapid and complete conversion of gold ion to $Au^0$, left an extremely low $C_{rGO}^{Au\ ion}$, consequently, a high $\Delta C$ was maintained during the adsorption process. Therefore, rGO overcame the fast equilibrium at low C, and showed an ultrahigh adsorption capacity at even ppb level. Secondly, the reductive adsorption was enabled by the graphene area of rGO, it donated electrons to the adsorbed gold ion which was reduced to $Au^0$. In addition, the wrinkles and warped area of graphene prompted the adsorption of gold ions and electron transfer to gold ions during the redox reaction, further increasing the gold extraction capacity. The oxidized regions of rGO, provided a good dispersibility of rGO, therefore allowing efficient adsorption and reduction of gold ions by graphene area, leading to an ultrahigh extraction capacity (Fig. 2h).

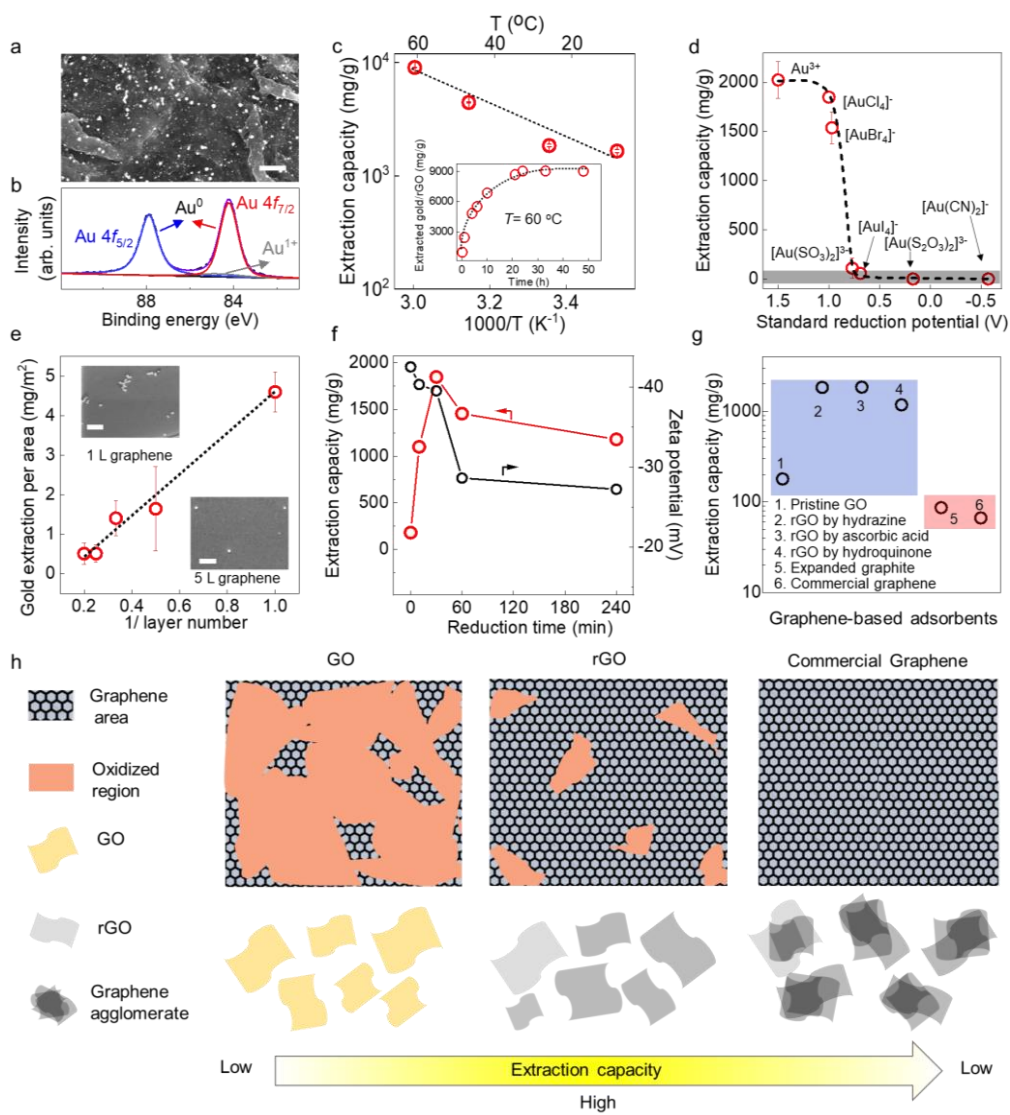

**Fig. 2| Understanding the gold extraction mechanism.** (**a**) Scanning electron microscopy (SEM) image and (**b**) 4$f$ XPS spectrum of gold nanoparticles on rGO. Black curve: raw data with the fitting envelope in purple. The deconvolved peaks for $Au^0$ and $Au^+$ are colour coded. Scale bar in (a), 500 nm. (**c**) Temperature dependent gold extraction behaviour. Inset is the extraction capacity as a function of time for a 10 ppm gold solution measured at $T$=60 °C. (**d**) Extraction capacity for gold complexes with different reduction potentials[42]. The shaded area marks extraction capacities below 100 mg/g. Dashed lines: guides to the eye. (**e**) Areal gold extraction capacity measured for mechanically exfoliated graphene. The error bars indicate standard deviation using several graphene crystallites. The insets are SEM images of exfoliated graphene ($N$= 1 and 5) after gold extraction. The scale bars are 500 nm. (**f**) Extraction capacity (measured at $C$= 10 ppm) and zeta potential of rGO versus its reduction time. (**g**) Extraction capacity measured for different graphene-based adsorbents. The light blue and light red shadings represent the graphene-based adsorbents with good dispersibility (blue), and those with well-retained graphitic area but poor dispersibility (red), respectively. (**h**) The schematic of atomic structures for GO, rGO, and commercial graphene, their corresponding dispersibility and extraction capacities. All the experiments except those in Fig. 2c were performed at 25 °C. All the error bars in this figure represent the standard deviation.

**Realization of high selectivity for gold extraction**

Based on the described understanding, we explore the possibility of extracting Au from e-waste that typically contains a variety of other metals. Initial tests were carried out using a simulated e-waste mixture containing $[AuCl_4]^-$, $Cu^{2+}$, $Ni^{2+}$, and $[PtCl_4]^{2-}$ ions. We observed ~99% gold recovery with <5 % of the other metals being extracted (Supplementary Fig. 13). Next, a discarded central processing unit (CPU) was leached with aqua regia, forming a leachate containing ~88 ppb of gold ions. Our rGO colloids captured ~90% of this gold (Fig. 3b) but also took up co-existing ions, similar to the other previously used gold adsorbents[4,10,12,14]. The uptake was from ~7 to 100 % for 6 out of the 14 metals present in the CPU leachate in concentrations from sub-ppm to hundreds of ppm (Fig. 3b).

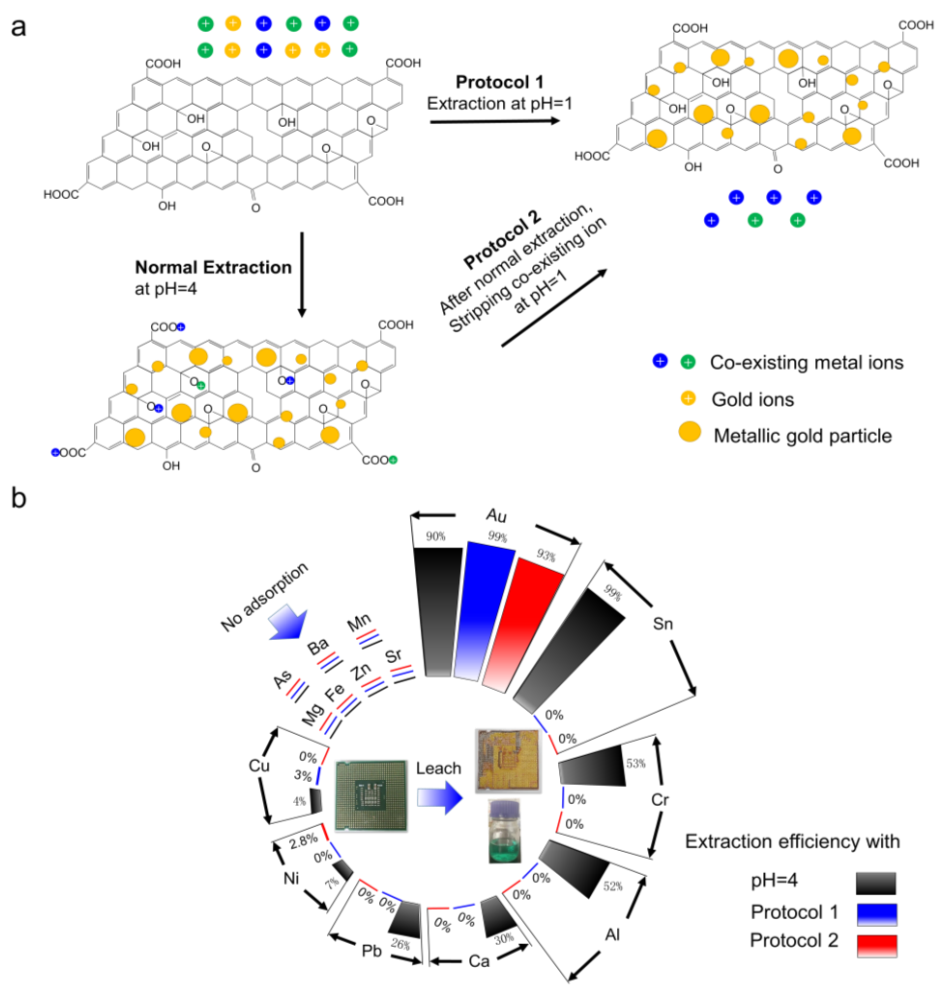

**Fig. 3| Highly selective gold extraction.** (**a**) Schematics show two additional protocols to increase selectivity by changing pH. Co-existing ions are prevented from adsorption (protocol 1) or stripped away later, after reductive adsorption of Au (protocol 2). (**b**) Comparison of ion selectivity using different extraction protocols. The photos inside the circle show the discarded, leached CPU and the resulting leachate. All the experiments were performed at 25 °C.

Because high selectivity is essential for viable recycling, we further improved our extraction procedures, considering that adsorption sites for gold and co-existing metal ions were likely to be different. Indeed, gold ions were reduced on graphene areas of rGO whereas residual functional groups were reported to provide efficient adsorption of metal ions[43-45]. By changing pH of rGO suspensions from basic to acidic, the functional groups could be (de-)protonated reversibly. Deprotonation results in more negatively charged rGO nanosheets that electrostatically attract and adsorb metal ions. In contrast, protonation prevents such adsorption. We exploited this consideration to achieve gold extraction with little co-adsorption of other ions within the CPU leachate (Fig. 3a). Two protocols were developed. In the first one, gold was extracted directly from a highly acidic (pH≈1) leachate (Fig. 3a). As shown in Fig. 3b, this allowed an extraction efficiency of 99.3% at gold $C \approx 88$ ppb whereas <3% of Al and Cu were extracted with undetectable adsorption of the other metals. This protocol could be particularly suitable for gold extraction involving strong acids, for example, if aqua regia is used to dissolve e-waste. In second protocol, we showed that the protonation is effective to strip the adsorbed co-existing ions on rGO after their adsorption. First, we exposed rGO to the CPU leachate at pH ≈ 4 for 24 h, which allowed maximum adsorption of gold. Then, pH was reduced to ~1 for an hour, which stripped off adsorbed co-existing ions (Fig. 3a). We obtained 93% gold extraction from the same CPU leachate with <2% for Ni and no other metals (Fig. 3b). Furthermore, for leachates with higher (2.65 ppm) Au concentrations and using the second protocol, noting in this case, the concentration of Cu was nearly two orders of magnitude higher than gold, still, ~99% of gold was extracted without detectable presence of any other metal (Supplementary Fig. 15). The success of these protocols strongly supports the suggested model of site-specific ion adsorption on rGO.

Gold extraction from seawater offers an extreme challenge because of minute Au concentrations being present, believed to be <20 ppt[12,46]. To address this challenge, we prepared simulated seawater and added 10 ppt of gold to it. Using our normal extraction procedures, we estimated complete gold extraction, indicating a possibility of gold mining from oceans (Supplementary Note 6). The Au uptake can probably be increased further by developing special protocols similar to those described above (Supplementary Fig. 14).

**Continuous flow-through gold extraction**

The described technology is scalable, as the process only involves mild temperature reduction, the rGO can be scalably made (Fig. 4a). Furthermore, to demonstrate its potential for scalable gold extraction, we also developed a scheme involving a continuous flow of low-$C$ Au solutions through rGO membranes (Fig. 4b; Supplementary Note 7). Fig. 4c shows the extraction performance of a membrane with a thickness of ~800 nm and 2 cm in diameter during filtration of several litres of a 100 ppb gold solution. As gold nanoparticles gradually accumulated between rGO nanosheets and started blocking the water flow (Fig. 4b), both extraction efficiency and membrane's permeance decreased. This decrease typical for any adsorption-based membrane separation[47,48] can be moderated using membranes in series. XRD analysis confirmed the presence of metallic gold within the resulting rGO membranes (Supplementary Fig. 16), their cross-sectional SEM and energy dispersive spectroscopy (EDS) imaging showed that gold was adsorbed and reduced through the entire cross-section (Fig. 4d). As an ultimate test for the technology's viability for e-waste recycling, the rGO membranes were used with real CPU leachates. The observed permeance and efficiency were close to those for pure gold solutions (Supplementary Note 8, Supplementary Fig. 17). As the final step to isolate metallic gold from our rGO, because of superior adsorption capacity and selectivity to gold, further elution and precipitation process are not required in our case. We therefore, used melting process which is the last step of gold recycling process, to burn the membranes in air at 700 °C, which left gold particulate (Fig. 4d). Its EDS analysis confirmed >95% purity with the rest being carbon, oxygen and sodium (Supplementary Fig. 16), this further proved that the adsorption of co-existing ions (both cationic and anionic ions) was little, in good agreement with the observed precise gold selectivity of rGO. We also tried regeneration of rGO by dissolving gold extracted by rGO with thiourea/HCl[16], and found regenerated rGO showed a gold extraction capacity of ~1,000 mg/g ($T$=25 °C). However, considering GO we used is a mass-produced commercial product which has a price of less than 0.5 RMB per gram (Shenzhen Matterene Technology), and high temperature used for removing rGO is, nevertheless, used for gold melting purpose, therefore, direct isolation at high temperature may be a better choice than regeneration with respect to the process cost and sustainability. It is worth noting that the nearly complete Au extraction from e-waste also simplifies the sortation of other metals present in leachates. For example, after gold extraction, we used the remaining filtrate to extract valuable copper by using its galvanic replacement with iron (Fig. 4d). The recycled copper had >95% purity (Supplementary Fig. 18).

Finally, we point out that the entire recycling process can be viable from an economic perspective. If we define the 50% extraction efficiency as the end life for rGO membranes, a 1 m$^2$ membrane (<3 g) under a 1 bar pressure is estimated to allow extraction of ~1.6 gram of gold from ~20 tons of a 100 ppb leachate. As the typical cost of GO is less than 0.5 RMB per gram, and the gold price is ~300 RMB per gram, together with the observed ultrahigh gold extraction capacity, precise selectivity, and scalability, though continuous efforts are required for its commercialization, rGO provides a considerable incentive for commercial recovery of gold from e-waste. Even rGO-based extraction from seawater might be viable as the process does not consume other reactants.

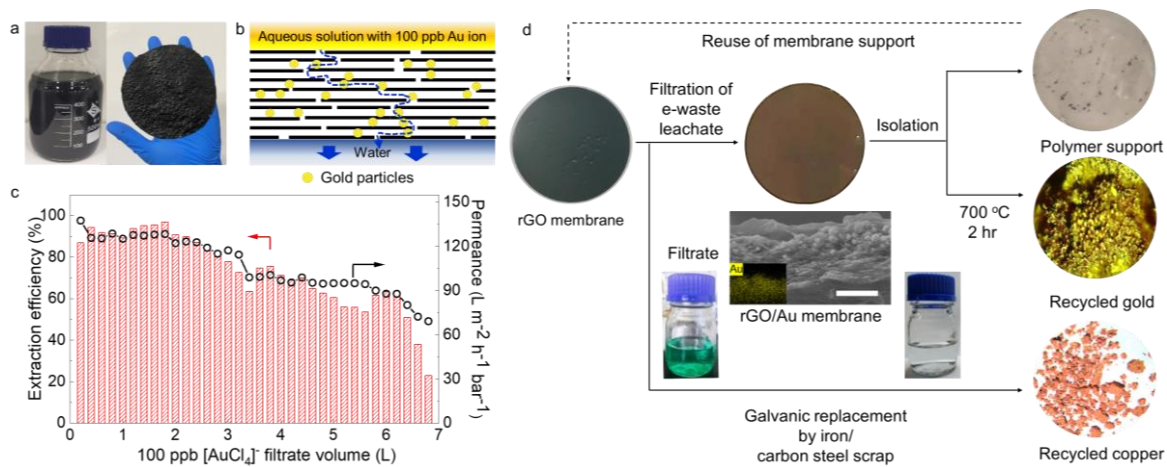

**Fig. 4| Flow-through technology for gold extraction and recycling. (a)** The left and right images show photos of rGO suspension and rGO membrane of 100 cm$^2$. **(b)** Schematic of low-concentration gold extraction using rGO membranes. **(c)** Extraction efficiency and water permeance reduced with increasing the amount of the 100 ppb Au solution filtrated through. **(d)** Schematic of a gold extraction process from e-waste leachates. The left, middle, and upper right images show photos of the initial rGO membrane, its state after filtration (at 50% permeance) and the polymer filter used for rGO membrane, respectively. Central panel: cross-sectional SEM micrograph of the final rGO membrane with gold nanoparticles seen accumulated between rGO nanosheets. The inset of the SEM image: EDS map of elemental gold. Scale bar, 500 nm. The container on the left is the leachate filtrate after gold extraction. The filtrate became colourless after galvanic extraction of copper (container to the right). The images presented as middle and lower circles to the right are photos of our final products: gold after burning rGO and galvanic copper, respectively. All the experiments were performed at 25 °C.

## Methods

**Preparation of rGO.** We used commercial GO (Shenzhen Matterene Technology) synthesized by a modified Hummers' method[49]. In a typical reduction process using ascorbic acid[50], pH of a 20 mL 0.5 mg/mL GO suspension was first adjusted to ~10 by adding an ammonia solution. Under stirring at room temperature, 35 mg of ascorbic acid was added, and the resulting GO suspension was placed for reduction in an oil bath at 95 °C for 30 minutes. This time was found optimal because shorter times diminished the area of unoxidized graphene whereas longer times resulted in agglomeration of rGO nanosheets, with both effects degrading the extraction capacity (Supplementary Fig. 10). The rGO suspension was centrifuged at 12,850 g, and the sediment was re-dispersed in water by sonication at 400 W for 30 min, resulting in rGO nanosheets with an average lateral size of 100-500 nm. This washing process was repeated at least 3 times to remove unreacted ascorbic acid and by-products generated during the reduction. The final rGO aqueous dispersion had a concentration of 0.5 mg/ml. For hydroquinone-reduced GO, 176 mg of hydroquinone was added to 20 mL 0.5 mg/mL GO suspension[51] and the same procedure as described above for ascorbic acid was employed. For making hydrazine-reduced GO, we followed the procedures reported elsewhere[52]. We mixed the 20 mL 0.5 mg/mL GO suspension, 50 μl of ammonia solution (28 wt% in water) and 14 μl of hydrazine solution (50 wt% in water). Then the same procedure as described above for ascorbic acid was carried out.

**Extraction capacity of rGO.** The capacity was measured using gold ion solutions ($KAuCl_4$) mixed with rGO suspensions to obtain mixtures with final gold concentrations of 0.1, 1, 10, 50 and 100 ppm. The gold solutions were prepared by diluting the stock solution of $KAuCl_4$ with deionized water. pH of the mixtures was adjusted to 4 by adding HCl or NaOH. The weight ratio of Au ion to rGO was kept at 2:1, and we note for 10 ppm and higher concentration, further increment of this weight ratio (Au ion to rGO) did not result in increment in extraction capacity at $T=25$ °C. We used extraction times from 2 minutes to 48 hours (Fig. 1c) while shaking the mixtures. At each time point, the mixtures were filtered using a 13 mm PES membrane syringe filter with 0.22 μm pore size. Then the filtrates were analysed by inductively coupled plasma optical emission spectroscopy (ICP-OES) and inductively coupled plasma mass spectrometry (ICP-MS) to determine the adsorption capacity, $Q_e$ (mg/g). It was calculated as $Q_e = \frac{(C_0 - C_e) \times V}{m}$ where $C_o$ is the initial concentration of Au (ppm), $C_e$ is its final concentration in the filtrate (ppm), $V$ is the volume of the used suspension (L) and $m$ is the mass of dry rGO (g). The gold extraction was usually performed in the dark, but no notable differences were found under light conditions. To measure the influence of pH, we followed the same procedures as above and changed pH by adding either HCl or NaOH. To ensure reproducibility, all the measurements were repeated at least 3 times. For gold extraction by mechanically exfoliated graphene is detailed in Supplementary Note 4, and for rGO's gold extraction from different gold complexes, we followed similar procedure as gold extraction from $[AuCl_4]^-$ solution, the exception was $[Au(CN)_2]^-$, that the extraction was performed at a pH=~10 as $[Au(CN)_2]^-$ becomes unstable in acidic solution. Unless otherwise noted, all experiments were conducted at room temperature (25 °C). In order to study the influence of temperature on gold adsorption properties, extraction experiments were conducted at 10, 25, 45, and 60 °C respectively. The weight ratio of Au ion to rGO was kept at 10:1 in this case.

**Extraction efficiency.** We measured extraction efficiency for 10 ppm solutions at pH of 2-11. The pH was adjusted by adding 0.5 M of HCl and 0.5 M of NaOH solutions. 10 mg of rGO were added to 20 mL of the gold solution with final gold concentration of 10 ppm. After shaking for 24 hours, the mixtures were filtered and analysed by ICP-OES to determine the extraction efficiency, $R$ (%). It was calculated as $R = \frac{C_0 - C_e}{C_0} \times 100\%$.

**Extraction selectivity.** In our single-metal selectivity tests, we used aqueous solution of $CuCl_2$, $NiCl_2$ and $K_2PtCl_4$ to achieve 10 ppm concentrations of $Cu^{2+}$, $Ni^{2+}$ and Pt (present as $[PtCl_4]^{2-}$ ions). The weight ratio of each metal ion and rGO was 2:1. For selectivity measurements in the presence of several metal ions, we mixed these solutions as well as a 10 ppm Au solution in equal volumes. Then, we followed the procedures described above for gold, repeating each experiment at least 3 times for reproducibility. All experiments were conducted at room temperature (25 °C).

**Real-world gold extraction.** A discarded CPU was obtained from computer waste. To leach gold from the CPU, it was first soaked in an 8 M NaOH solution for two days to remove the protective coating on the electronic surfaces. After that, the CPU was rinsed and soaked in 40 mL aqua regia at 60 °C for two days. Undissolved material was filtered and rinsed with deionized water to form a leachate with $C$ ~100 ± 20 ppb. pH of the leachate was changed to 4 by adding NaOH. 3.6 mg of rGO was added to 20 mL of the final leachate and stirred for 24 h at room temperature. Then we followed the procedures described above for gold extraction measurements. In the first protocol, we adjusted the pH value of the leachate to 1, using HCl and only then performed the gold extraction. In the second protocol, after gold adsorption on rGO at pH ≈ 4, we again used HCl to reach pH ≈ 1, and the mixture was shaken for an extra hour. The amount of extracted gold was evaluated using the procedures described above with at least 3 samples. All experiments were conducted at room temperature (25 °C).

**Flow-through extraction and copper recycling.** A suspension containing ~1.8 mg of dry rGO was filtered through a cellulose membrane support for 12 h by vacuum filtration at 25 °C. Note that, if we vacuum-filtrated for much longer times (e.g., 72 h), rGO deposits became completely dry and exhibited no permeability to water. The resulting (wet) rGO membranes were tested for their performance using a 100 ppb gold solution. The permeance and gold extraction efficiency were measured each time after 200 mL of the solution was filtered. The efficiency was calculated as $R = \frac{C_0 - C_e}{C_0} \times 100\%$. We found a trade-off between the permeance and efficiency such that thicker membranes led to a lower permeance but higher uptake of gold (Supplementary Fig. 14). We chose to work with 800 nm-thick membranes (measured in the dry state) because of a good balance between permeance and uptake.

To demonstrate copper extraction from e-waste, we used the leachate left after extracting gold by filtration through rGO. Iron particles were added directly into the filtrate for 30 minutes after which no further precipitation was noticeable. The reddish-brown precipitate was then collected, washed in 0.1 M HCl and then in deionised water, and vacuum dried for chemical analysis.

**Materials Characterization.** Metal concentrations in aqueous solutions were measured using ICP-MS (iCAP RQ) and ICP-OES (Optima 7300 DV). XRD was performed using a Bruker D8 advance diffractometer operated at 40 kV and 40 mA using Cu Kα radiation ($\lambda$ = 0.154 nm). The morphology of rGO after extraction was examined by SEM (Hitachi SU8010) operated at 5 kV. DSC and TG measurements of KAuCl$_4$, rGO and the products after gold extraction were obtained using analyser Jupiter STA 449 F3 at heating and cooling rates of 10 °C/min from 20 to 800 °C. XPS measurements were conducted on spectrometer PHI-5000 Versa Probe II. Zeta potential measurements were performed using Zeta sizer Nano ZS. UV–Vis absorption spectra were recorded on a UV−Vis spectrophotometer (UV-2600, Shimadzu). Raman spectra were recorded on a Horiba Evolution HR Raman spectroscope using a 532 nm argon ion laser.

**Data availability**

The data that support the plots within this paper and other findings of this study are available within the article and the Supplementary Information file, or available from the corresponding authors upon request. Source data are provided with this paper.

**Highly Efficient and Selective Extraction of Gold by Reduced Graphene Oxide**

Fei Li[#,1], Jiuyi Zhu[#,1], Pengzhan Sun[2], Mingrui Zhang[1], Zhenqing Li[1], Dingxin Xu[1], Xinyu Gong[1], Xiaolong Zou[1], A. K. Geim[1,2]*, Yang Su[1,]*, Hui-Ming Cheng[3,4,]*

#1 Comparing rGO's performance with other gold adsorbents

Figure S1 compares our rGO suspensions with gold adsorbents reported in the literature[1-21]. A large amount of published data is summarized in this figure, which clearly shows a trade-off between the extraction capacity and gold concentration $C$ in solutions.

To highlight the superior extraction capacity of rGO, we compared rGO's extraction capacity with adsorbents that showed high extraction capacity reported elsewhere. The molybdenum disulfide modified carbon nanotubes (CNT-MoS$_2$), thiourea-modified porous aromatic framework (PAF-1-thiourea), porous porphyrin polymer (COP-180), and amyloid-like protein membrane (PTL membrane) showed a maximum capacity of 2495 mg/g to 1000 ppm Au ion, 2629 mg/g to 500 ppm Au ion, 1620 mg/g to 3000 ppm Au ion, 1034 mg/g to 984.8 ppm Au ion, respectively. However, their capacity decreased dramatically to low concentration. For example, the extraction capacities to the lowest gold concentration studied in these reports are 1000 mg/g to 100 ppm Au ion (CNT-MoS$_2$), 250 mg/g to 20 ppm Au ion (PAF-1-thiourea), 100 mg/g to 20 ppm Au ion (COP-180) and 500 mg/g to 196.9 ppm Au ion (PTL membrane), respectively. In comparison, rGO showed a significantly higher capacity as shown in Fig. S1a and Fig. S1b. Moreover, this exceptional performance of rGO suspensions extends into the ppb and sub-ppb range where no other adsorbent was so far reported to exhibit any discernible extraction of gold (Fig. S1c).

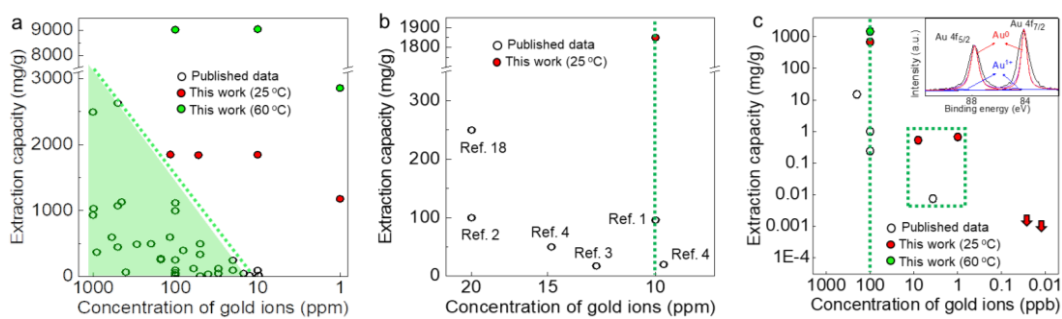

**Fig. S1| Comparison of gold extraction performance by rGO with other adsorbents.** (**a**) Extraction capacity as a function of gold concentration above 1 ppm. Open symbols are the data taken from Refs. 1-21 and listed in Table 1. (**b**) Extraction capacity as a function of gold concentration range from 10 to 20 ppm. (**c**) Same for concentrations below 1 ppm. The red arrows show the estimated range of extraction capacity at 10 ppt and 20 ppt, we estimate ~100% extraction efficiency based on X-ray photoelectron spectroscopy (XPS) analysis. Note the logarithmic scale for the y-axis. Inset is 4f XPS spectrum of gold extracted by rGO from 1 ppb [AuCl$_4$]$^-$ measured by XPS. The open symbols are the data taken from Refs. 1 and 22-23. All the data compared in Fig. S1a was listed in Table S1.

For determination of extraction efficiency and capacity to 1 ppb, 20 ppt and 10 ppt gold solution, 0.3 mg rGO was added in 200 mL of gold solution with above concentrations. For 1 ppb gold solution, we obtained ~100 % extraction efficiency measured by inductively coupled plasma mass spectrometry (ICP-MS), which translated into an extraction capacity of ~0.7 mg/g, and similar to higher concentration, XPS analysis showed extracted gold was predominately $Au^0$ (Fig. S1c inset). For rGO's adsorption to 20 ppt and 10 ppt gold solution, as ICP-MS approaches its detection limit, the rGO nanosheets after extraction were collected by centrifugation and drop-casted on a Si wafer substrate for XPS analysis. The gold contents that measured at different areas (at least 5 areas) varied from 0.01-0.18 at%. Thus, as a semi-quantitative method, XPS clearly validates significant gold extraction by rGO at ppt level.

**Table S1| Extraction capacities of various gold adsorbents.** Note that all of the gold complex ion form in different references was the $[AuCl_4]^-$ or $Au^{3+}$, except for the $Au(S_2O_3)_2^{3-}$ in the literature 4 and 20.

| gold concentration (ppm) | Types of gold ion | Gold adsorbents | Extraction capacity (mg/g) | Reference |
|---|---|---|---|---|
| 10 | | Fe-BTC/PpPDA | 96 | 1 |
| 1000 | | Fe-BTC/PpPDA | 934 | 1 |
| 20 | | COP-180 | 100 | 2 |
| 3000 | $[AuCl_4]^-$ | COP-180 | 1620 | 2 |
| 12.5 | | TDAC | 17.5 | 3 |
| 100 | | TDAC | 30 | 3 |
| 14.8 | | $MoS_2$/CS aerogel | 50 | 4 |
| 116 | $Au(S_2O_3)_2^{3-}$ | $MoS_2$/CS aerogel | 600 | 4 |
| 9.7 | | $MoS_2$/CS aerogel | 20 | 4 |
| 30 | | COP-224 | 50 | 5 |
| 150 | | UiO-66-TA | 260 | 6 |
| 900 | $[AuCl_4]^-$ | UiO-66-TA | 372 | 6 |
| 80 | | CSGO5 | 400 | 7 |
| 500 | | CSGO5 | 1076 | 7 |
| 591 | $Au^{3+}$ | $CaCu_6[(S,S)$-methox$]_3(OH)_2(H_2O)$ | 598 | 8 |
| 100 | | UiO-66 | 60 | 9 |
| 100 | $[AuCl_4]^-$ | UiO-66-$NH_2$ | 100 | 9 |
| 150 | | UiO-66-TU | 275 | 10 |
| 30 | | SH-MCM-41 | 125 | 11 |
| 100 | | barley straw carbon | 256 | 12 |
| gold concentration (ppm) | Types of gold ion | Gold adsorbents | Extraction capacity (mg/g) | Reference |
| 293 | [AuCl$_4$]$^-$ | barley straw carbon | 492 | 12 |
| 50 | | Fe$_3$O$_4$@DMSA | 340 | 13 |
| 50 | | L-lysine modified, crosslinked chitosan resin | 13 | 14 |
| 400 | | L-lysine modified, crosslinked chitosan resin | 70.34 | 14 |
| 40 | | cross-linked lignocatechol | 40 | 15 |
| 60 | Au$^{3+}$ | modified wheat straw | 125 | 16 |
| 500 | | modified wheat straw | 450 | 16 |
| 1000 | [AuCl$_4$]$^-$ | CNT-MoS$_2$ | 2495 | 17 |
| 100 | | CNT-MoS$_2$ | 1000 | 17 |
| 500 | | PAF-1-thiourea | 2629 | 18 |
| 20 | | PAF-1-thiourea | 250 | 18 |
| 984.8 | Au$^{3+}$ | PTL membrane | 1034 | 19 |
| 196.9 | | PTL membrane | 500 | 19 |
| 100 | Au(S$_2$O$_3$)$_2^{3-}$ | MoS$_2$/ZnS | 1120 | 20 |
| 50 | | MoS$_2$/ZnS | 500 | 20 |
| 450 | Au$^{3+}$ | MoS$_2$ | 1133 | 21 |
| 100 | [AuCl$_4$]$^-$ | rGO nanosheets at 60 °C | 9034 | this work |
| 10 | | | 9059 | |
| 1 | | | 2858 | |
| 0.1 | | | 1480 | |
| 100 | | rGO nanosheets at 25 °C | 1880 | |
| 10 | | | 1850 | |
| 1 | | | 1180 | |
| 0.1 | | | 690 | |

Fig. S2 showed the change of gold extraction capacity with pH at a temperature ($T$) of 25 °C (Fig. S2a) and 60 °C (Fig. S2b). At $T$=25 °C, rGO had extraction capacities of 700 mg/g and 340 mg/g to 10 ppm gold at pH=2 and 11 respectively. The extraction capacity can be further increased by increasing $T$. At 60 °C, rGO had a capacity of 3200 mg/g at pH=2, even at pH=0 (10 ppm Au solution containing ~2 M H$^+$), we still observed a capacity of 586 mg/g. At pH=11, the extraction capacity of rGO to 10 ppm gold was 1565 mg/g. In contrast, as shown in Fig. S1 and table S1, most adsorbents showed an extraction capacity <300 mg/g at 10 ppm, suggesting superior gold extraction performance of rGO even in strong acidic and basic solutions.

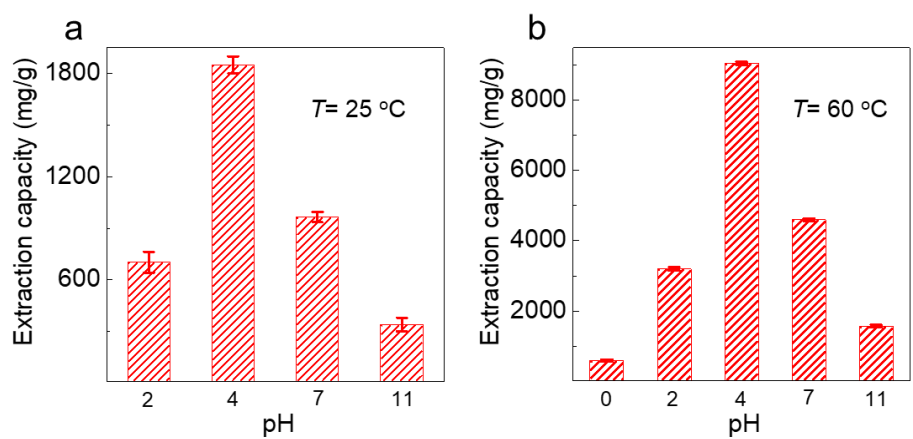

**Fig. S2| The extraction capacity measured for 10 ppm solutions at different pH**. The weight ratio of gold ion and rGO was 2:1. The extraction temperature was (a) 25 °C and (b) 60 °C.

To study the gold adsorption isotherms of rGO, batch adsorption experiments were conducted. 2 mL rGO suspension (0.5 mg/mL) was added to 198 mL gold solution with initial concentrations of 1, 5, 10, 20, 50 and 100 ppm, respectively. After 24 hr adsorption, the extraction capacity was measured and plotted versus the equilibrium gold concentration (Fig. S3a). At an equilibrium concentration of 0.21 ppm (10 ppm as starting concentration), the adsorption capacity reached 1850 mg/g, confirming ultrahigh gold extraction capacity. By normalizing the extraction capacity at different time, it became more evident to support the fast adsorption kinetics. Within 10 minutes, its extraction capacity was 1012 mg/g and extraction efficiency reached 54.7%, and reached 100% after 24 hr (Fig. S3b).

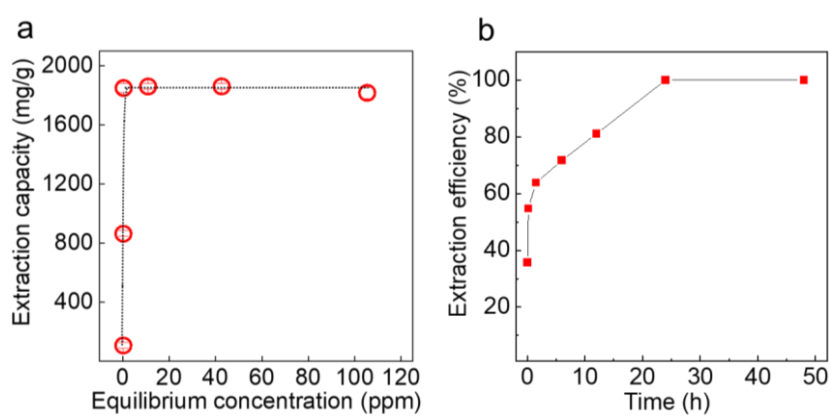

**Fig. S3| Gold adsorption isotherms and kinetics at pH=4**. **The extraction temperature was 25 °C.** (a) Gold adsorption isotherms. (b) Gold adsorption kinetics normalized with full capacity at pH=4. The weight ratio of gold ion and rGO was 2:1.

#2 Characterization of graphene oxide and rGO before and after Au extraction

To characterize our graphene-based materials, we used X-ray diffraction (XRD) combined with other analytical techniques, including XPS and Fourier transform infrared (FTIR) spectroscopy (Nicolet iS50).

The XRD patterns of GO, rGO before and rGO after Au extraction (denoted below as rGO-Au, 10 ppm [AuCl$_4$]$^-$ was used) are shown in Fig. S4a. The XRD confirmed that GO was reduced by ascorbic acid. Indeed, the ~10° peak known for GO disappeared whereas another peak characteristic of rGO emerged at ~23°. The gold on rGO-Au was found to be metallic as determined from its XRD peaks (XRD peak was fitted according to JCPDS data, No. 04-0784).

The FTIR spectra of GO, rGO, and rGO-Au are shown in Fig. S4b. The peaks at 3431, 1725 and 1396 cm$^{-1}$ are assigned to the O-H stretching vibrations, C=O stretching and O-H flexural vibration modes of carboxylic groups, respectively[24]. In rGO and rGO-Au samples, the intensities of peaks at 3431 cm$^{-1}$ and 1725 cm$^{-1}$ decreased notably compared to the GO, indicating the removal of functional groups. The peak at 1396 cm$^{-1}$ is still visible in both rGO spectra, yielding the presence of residual oxygen-containing functional groups that contribute to the stability of our rGO dispersions. The FTIR spectrum of rGO-Au exhibited no notable changes as compared to rGO, which is consistent with our interpretation that gold ions are adsorbed mostly onto graphitic regions of rGO (see the main text).

To validate gold ion has been reduced to metallic gold during extraction, thermogravimetric (TG) and differential scanning calorimetry (DSC) were used to analyse rGO sample after 24 hours extraction (rGO-Au-24 h). Fig. S4c shows the TG and DSC curves of rGO and KAuCl$_4$. The weight losses of rGO and KAuCl$_4$ during heating were ascribed to the decomposition of rGO and [AuCl$_4$]$^-$ → Au$^0$, respectively. For KAuCl$_4$, such transformation led to an endothermic peak at ~330 °C, which is in agreement with the previous report[25]. In contrast, this peak was absent for rGO-Au-24 h (Fig. S4d), suggesting the extracted gold in rGO-Au-24 h was mainly Au$^0$ other than [AuCl$_4$]$^-$. Furthermore, after 700 °C calcination, the rGO-Au-24 h remained ~69 wt% of its original weight (Fig. S4d), giving an extraction capacity of ~2100 mg/g (considering 3.6 wt% remained ash for pristine rGO (Fig. S4c), in good agreement with the extraction capacity measured by ICP-MS (Fig. 1b in the main text).

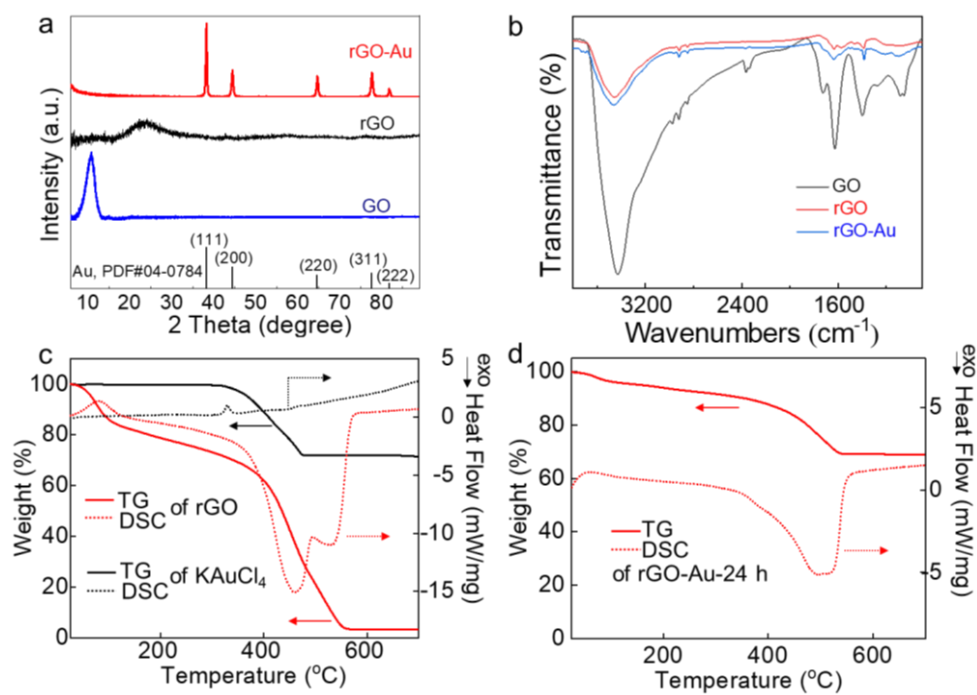

**Fig. S4| Characterization of GO, rGO, and rGO-Au.** (**a**) Their XRD patterns and (**b**) FTIR analyses. (**c**) TG and differential scanning calorimetry (DSC) curves of $KAuCl_4$, rGO, and (**d**) rGO-Au after 24 h extraction (rGO-Au-24 h) measured in air.

The Raman spectra in Fig. S5a show that the $I_D/I_G$ ratios of rGO and rGO-Au are 1.06 and 0.94, respectively. The defect density in the carbon materials, especially graphene materials is characterized by $L_D$, the distance between two neighboring defects. Obviously, a higher $L_D$ suggests a less defective carbon material[26]. $L_D$ could be determined from $I_D/I_G$, and generally does not monotonically change with $I_D/I_G$. At a small $L_D$, an increase in $I_D/I_G$ suggests an increased $L_D$. After reaching a maximum, $L_D$ further increases with decreased $I_D/I_G$. GO and rGO are reported with a small $L_D$ that increases with an increased $I_D/I_G$. GO and rGO are reported with a small $L_D$ that decreases with a decreased $I_D/I_G$, therefore, the observed decrement of $I_D/I_G$ after gold extraction suggested a more defective state of rGO, supporting the electron donation from rGO to gold. In addition, G band of rGO after gold extraction showed a blueshift from 1602 to 1606 $cm^{-1}$, confirming a p-doping and electron transfer from rGO to gold[27].

Fig. S6 are the Raman map of G and D bands of rGO before and after the gold extraction at 25 °C and 60 °C respectively. We found that, firstly, because of the existence of gold on the rGO surface, the surface enhanced Raman scattering (SERS) effect emerged. Specifically, $I_D$ peak showed an intensity range from 300-1400 for rGO, increased to 1500-2600 after gold extraction, and $I_G$ peak increased from 300-1200 to 1400-2200. Secondly, such mapping allowed us to summarize the change of $I_D/I_G$ before and after gold extraction. In good agreement with Fig. S6a, $I_D/I_G$ decreased from a range of 1.00-1.15 to 0.93-1.02 after extraction, suggesting a more defective rGO after gold extraction, because of the electron donation.

Fig. S5b shows the UV-Vis spectra of GO, rGO, and rGO-Au. Compared to GO, the peak at 227 nm corresponding to $\pi \rightarrow \pi^*$ transitions of aromatic C–C bonds shifts to 263 nm after reduction, indicating the restoration of the electronic conjugation within the graphene sheets[28]. After mixing the rGO with $[AuCl_4]^-$, i.e. sample rGO-Au, we observed a blueshift of rGO characteristic absorption peak from 263 nm to 232 nm after 24 hr extraction, indicating electron transfer from the graphitic area to $[AuCl_4]^-$ and reducing $[AuCl_4]^-$ to $Au^0$, the adsorption peak of $Au^0$ located at 555 nm, which is typical for gold nanoparticles.

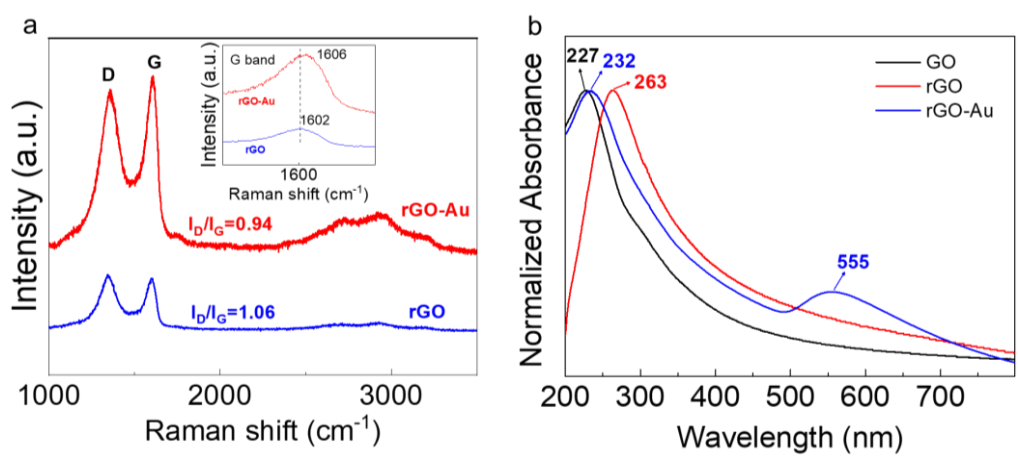

**Fig. S5| Evolution of rGO during the gold extraction process.** (a) Raman spectra of the rGO and rGO-Au. The inset shows the blueshift of G peak from 1602 to 1606 cm$^{-1}$ after gold extraction. (b) UV-Vis spectra of GO, rGO, and rGO-Au.

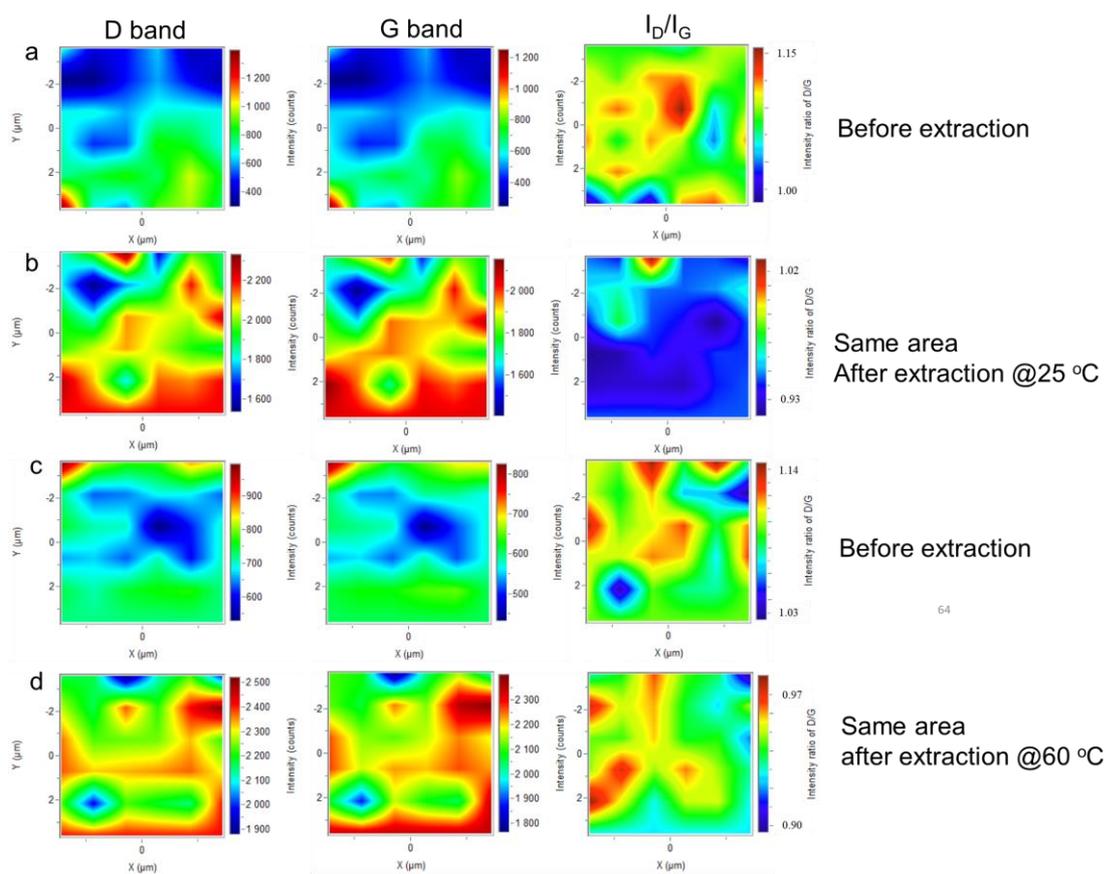

**Fig. S6| Raman map images for D, G and D/G of rGO before and after gold extraction at 25 °C and 60 °C**: (a) D band, G band, and $I_D/I_G$ before gold extraction at 25 °C. (b) D band, G band, and $I_D/I_G$ after gold extraction at 25 °C. (c) D band, G band, and $I_D/I_G$ before gold extraction at 60 °C. (d) D band, G band, and $I_D/I_G$ after gold extraction at 60 °C. Note that each band was scanned in the same area of 7.2 × 7.2 μm².

#3 Changes in rGO during gold extraction

SEM was employed to observe changes in rGO during the gold extraction process (Fig. S7). It can be seen that, even after 2 minutes, gold nanoparticles already appeared, suggesting that the reductive adsorption mechanism kicked in. It was clear that each gold particle has an intra-particle distance between tens nanometers to a few hundred nanometers, this suggested the electron transfer needed for reductive adsorption may only require electron transfer in the sub-micrometre range, so that the interconnected graphene areas of rGO were able to provide electrons and reduce gold ion at its vicinity.
.

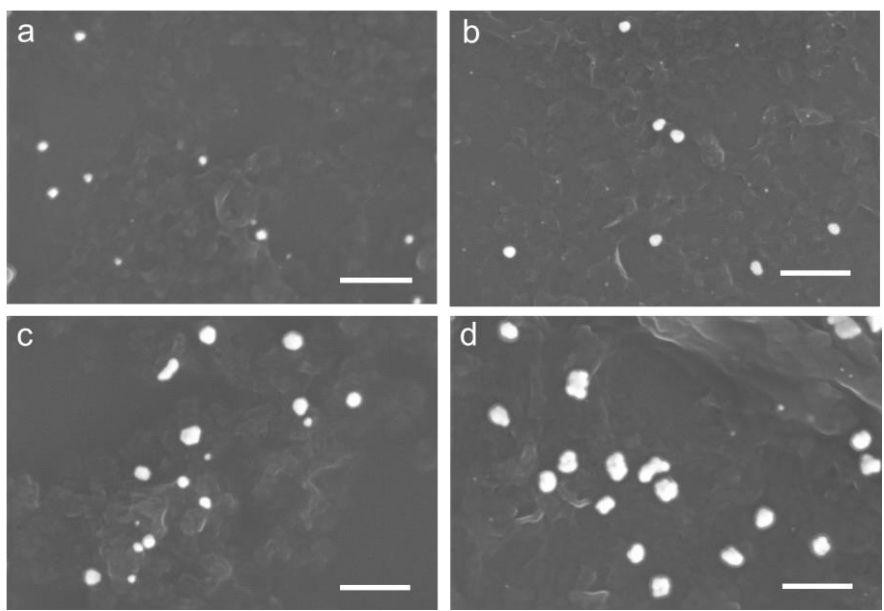

**Fig. S7|** SEM images of rGO nanosheets after their exposure to a 10 ppm gold solution of KAuCl$_4$ for 2 min, 10 min, 1.5 h, and 12 h. Panels (a), (b), (c), and (d) respectively. All scale bars, 200 nm.

To check whether it is metallic gold or KAuCl$_4$ salt adsorption dominated in the early stages, we analysed rGO-Au after 10 minutes extraction (rGO-Au-10 min, 10 ppm gold solution is used) by TG and DSC analysis (Fig. S8). Similar to rGO-Au-24 h, we did not observe the corresponding peak for [AuCl$_4$]$^-$ →Au$^0$ for rGO-Au-10 min. This is similar to the behaviour found for rGO-Au after 24 h reduction (Fig. S4d), and confirms that the reductive adsorption mechanism takes place rapidly, at least within less than a few minutes. In addition, TG showed ~56.7 wt% for rGO-Au-10 min, which gives an extraction capacity ~1.2 g/g for 10 min gold extraction, in good agreement with the capacity measured by ICP-MS (Fig. 1c in the main text).

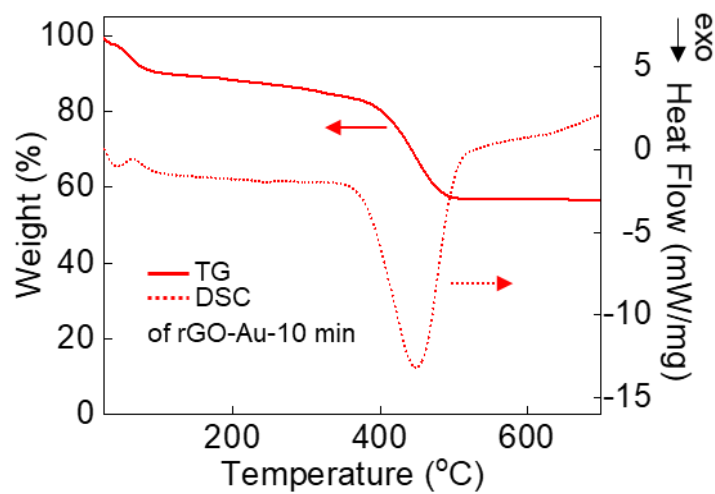

**Fig. S8|** TG and DSC curves of KAuCl$_4$ and rGO-Au-10 min measured in air.

#4 Gold reduction on mechanically exfoliated graphene

To gain further insight into the Au reduction mechanism, we studied the influence of graphene's thickness and morphology on its gold extraction ability. To this end, we prepared pristine graphene crystals by the standard exfoliation technique on top of an oxidized Si wafer[29]. Their thickness (the number of layers, $N$) was identified using optical contrast[29]. The obtained crystals were then exposed to a 10 ppm $KAuCl_4$ aqueous solution. Figure S9 exemplifies our observations. The SEM image shows a region covered with mono- and bi- layer graphene. After its exposure to the Au solution for 5 minutes, many areas of monolayer graphene became scrolled, warped and folded. These structural distortions were also observed in bilayer and few-layer regions (Fig. S9) whereas multilayer graphene and graphite crystals remained flat (not shown). These pronounced changes in morphology are attributed to water permeating under exfoliated crystals, which weakened their adhesion to the $SiO_2$ substrate allowing the folding. We note such morphology changes did not require any presence of Au salts and occurred in both pure water and Au salt solutions before drying samples. For example, these changes in morphology of graphene crystals were observed in an optical microscope in situ, inside deionized water.

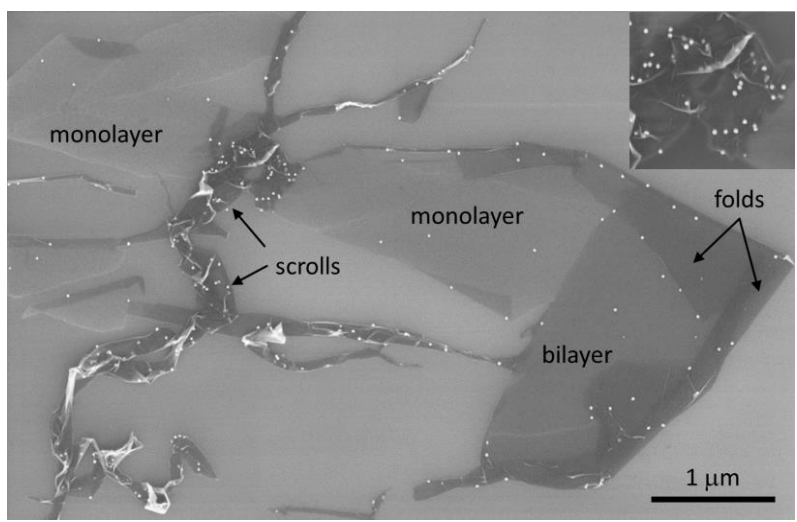

**Fig. S9| Gold extraction on flat and warped areas.** Micrograph of graphene after its exposure to a 10 ppm Au ion solution for 5 minutes. Gold nanoparticles (seen as white dots and confirmed by EDS) were mostly found on top of scrolls and wrinkles, and the nanoparticles also decorated edges of folded areas. Inset: Zoom-in of the warped graphene area in the centre of the top-left quadrant. The SEM images were taken at 5 kV using electron microscope ULTRA by Zeiss.

Fig. S10 shows that Au nanoparticles (~ 20 nm in diameter) were found mostly on top of warped and folded regions with only a few particles present in flat areas. By analysing the resulting surface coverage using many such SEM images, we evaluated that the flat areas contained ~0.1 mg of Au per m$^2$ and that there was no statistically significant difference in the amount of gold deposited onto flat regions of different thicknesses (we analysed areas with $N$ from 1 to 5 and thick graphite). The above amount translates into the extraction capacity $M$ of ~150 mg of Au per g of monolayer graphene and progressively less for thicker crystals ($M \propto 1/N$). It is difficult to evaluate accurately $M$ for warped areas because of their unknown and varying thickness. However, the SEM contrast suggests that they contained graphene monolayers (and occasionally bilayers) folded only a few times. This allows us to estimate $M$ for warped graphene as ~ 1,000 mg per g of carbon, in agreement with the extraction capacity observed for rGO for short times (Fig. 1c in the main text). In this respect, it is important to emphasize that rGO also consists of scrolled, wrinkled and folded areas rather than flat graphene.

Next, similar samples containing graphene and graphite were exposed to the same 10 ppm Au solution for 19 hr. Again, Au nanoparticles were found to heavily cover warped areas, but the coverage of flat areas was also denser allowing statistical analysis. The results are summarized in Fig. S11 and Fig. 2e. First, Au nanoparticles became noticeably bigger for all $N$ and occasionally could reach up to 100 nm in size. The particles also acquired irregular shapes as shown in Fig. S10, suggesting a merger of several smaller particles. The areal Au extraction was found to be highest for monolayers, decaying with increasing $N$ but recovering to mid values for thick graphite crystals. This behaviour is illustrated by micrographs of Fig. S10 and quantified in Fig. 2e. We attribute the higher coverage observed for $N$ = 1 to the fact that visibly flat areas of monolayer graphene were not atomically flat but followed the morphology of the oxidized Si wafer. Ripples on graphene were previously shown to be catalytically active[30]. Accordingly, the high coverage of monolayers could be due to the same effect as seen in Fig. S9 for warped graphene. For larger $N$, crystals became increasingly flat, leading to fewer Au nanoparticles. It remains to be understood why graphite surfaces also contained a reasonably high Au coverage, higher than that on few-layer graphene. To this end, we note that our graphite crystals contained cleavage steps and some folded areas. This allows us to speculate that Au reduction occurred predominantly on the steps and folds (catalytically active features) and then nanoparticles migrated along the atomically flat surfaces of graphite crystals, leading to their relatively uniform coverage. Results of Fig. S10 and Fig. 2e yield the extraction capacity $M$ for monolayer graphene of ~ 6,000 mg per gram of graphene, that is, ~3 times higher than for rGO in Fig. 1 of the main text. Such enhancement is perhaps not surprising because rGO nanosheets 1) contain oxidized areas (that is, not the entire surface could take part in reduction and it might contribute to the energy barrier for gold reduction as observed in Fig. 2c) and 2) tend to coagulate after initial stages of gold adsorption so that some of the graphene areas become inaccessible to gold ions.

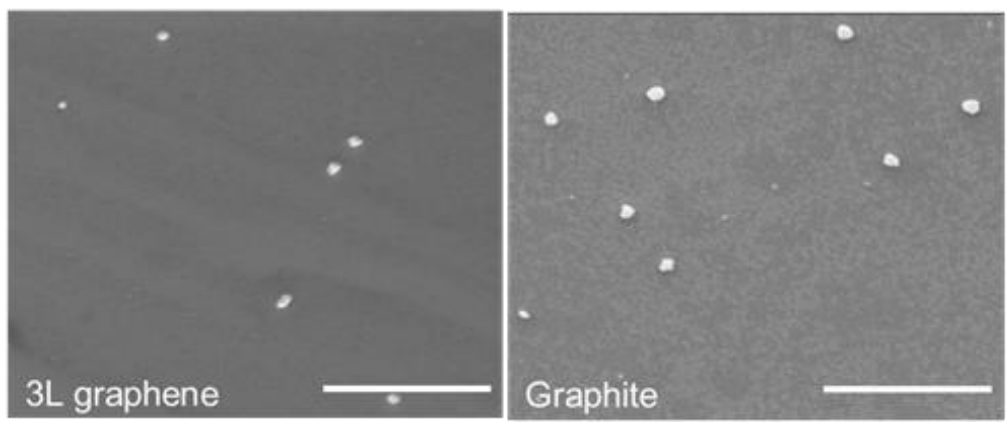

**Fig. S10| Gold extraction for graphene of different thicknesses.** Micrographs illustrating typical Au coverage for flat areas of graphene with $N$ = 3 and graphite. Note, N=1 and 5 are presented as insets in Fig. 2c in main text. Exposure: 19 hours in a 10 ppm Au solution. Scale bars, 1 μm for all images.

To gain the understanding on why warped area of graphene enhance gold extraction behavior, we studied adsorption energy and charge transfer between gold ion and graphene using First-principle calculation (Fig. S11). All calculations were carried out using the Vienna Ab initio Simulation Package (VASP)[31,32] based on density functional theory (DFT)[33,34] with the Perdew-Burke-Ernzerhof functional[35]. As shown in Fig. S11a and b, AuCl$_3$ cluster was adopted to represent the valence state of Au$^{3+}$, while (10, 10) carbon nanotube (CNT) was chosen as a similarity of the warped and curved surface of graphene. 6×6×1 and 1×1×5 supercells for graphene and nanotube were constructed to study the adsorption of Au$^{3+}$. The vacuum layers were set as at least 12 Å to avoid spurious interactions among periodic images. Zero damping DFT-D3 method[36] was applied to describe van der Waals interaction. The adsorption energy was defined as $E=E_T-E_C-E_{AuCl3}$, where $E_T$, $E_C$ and $E_{AuCl3}$ are the total energies of the adsorption system, graphene or nanotube supercell, and AuCl$_3$ cluster, respectively.

Our results show that, in contrast to flat graphene, the adsorption energy of gold ion on curved graphene surface is about 0.1 eV lower than that on graphene, indicating its preferred adsorption on the curved graphene surface. After adsorption, the electron transfer process from graphene to gold ion drives the reduction of gold ion to Au$^0$. Our calculation shows that, the curved graphene has a Fermi level ~0.3 eV higher than that of flat graphene, this leads to a more significant charge transfer from curved graphene. This is further validated by Bader charge analysis, that we found that the numbers of electrons transferred from curved and flat graphene to gold ion are 0.61 and 0.47, respectively, in good agreement with observed significant gold reductive adsorption on curved surfaces.

To further determine the difference in electron transfer, we also performed gas-phase calculations on cluster models, which can take different charge states. Fig. S11c and d shows the structural models for Au-adsorbed graphene and CNT clusters, which were saturated with H atoms. The +3 charge state of the cluster models was realized by artificially setting the number of electrons, and the compensating background charge was included to ensure the convergence of electrostatic energy. After structural relaxation, the Bader charge analysis shows that the number of electrons transferred from CNT and graphene clusters to $Au^{3+}$ was 3.03 and 2.81, respectively. Compared to graphene, CNT transferred 0.22 more electrons to $Au^{3+}$, which is consistent with the calculation based on $AuCl_3$ models (0.13 e). Here, we used the idealized charged systems, and a jellium background charge was added to ensure the whole system is neutral. It should be noted that, the jellium charge might introduce spurious states in the vacuum[37], which could result in the change to calculated one-electron energy and charge distribution and thus the Bader charge. Nevertheless, our DFT calculations showed a consistent trend with the experimental results. To obtain more accurate results, a self-consistent correction[37] or gas-phase simulations should be explored.

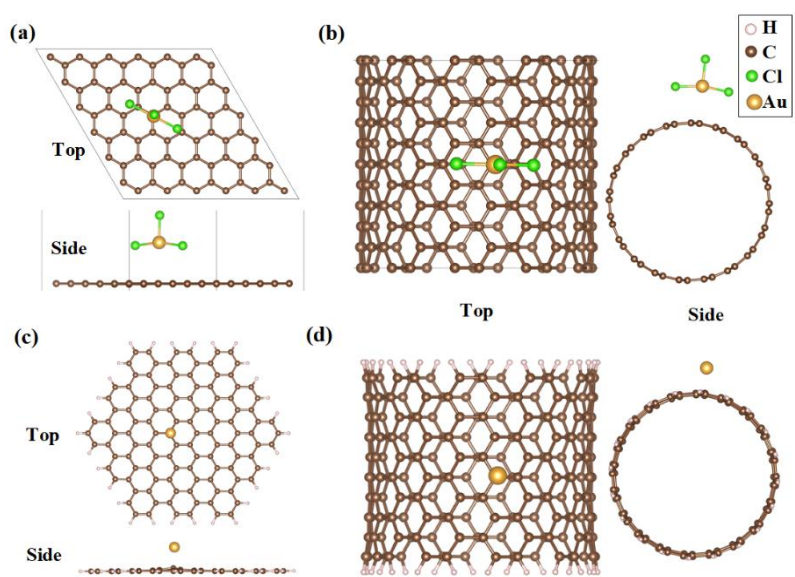

**Fig. S11| The models of gold ion adsorption on (a) graphene and (b) CNT.** Gold ion is placed on top of carbon atom as it is the most stable configuration of adsorption. Top and side views of Au-adsorbed (c) graphene and (d) CNT clusters.

To summarize, our observations for exfoliated graphene suggest that both surface area and thickness are important for efficient Au extraction. The available area is obviously maximal for monolayer graphene, as in rGO's solutions and membranes used in our work. The extraction capacity is negligible for graphite (in terms of gold extracted per gram) and rapidly increases as 1/*N* with decreasing the graphite thickness *N*. Moreover, monolayer graphene is also beneficial for Au reduction by speeding up the process on top of uneven and warped areas, which are abundant within rGO nanosheets. In addition, our experiments on exfoliated graphene confirmed that it was pristine graphene areas that were important for Au reduction, and the minority rGO areas that remain oxidized played little role in the process.

#5 Evaluation of different graphene-based adsorbents

For the rGO reduced by ascorbic acid with different time, we have used XPS, specifically, changes in C/O ratio of resulting rGO, to confirm that control of reduction time can tune the oxidized region of rGO. As shown in Fig. S12, pristine GO showed a C/O ratio of 2.2, which has increased to 4.2, 4.7, 5.1, 5.7 for a reduction time of 10 min, 30 min, 1 h, 4 h respectively, as discussed in the previous paper, such increase of C/O ratio was strong evidence for the removal of oxidized region[24,38].

For the rGO reduced by hydrazine and hydroquinone, commercial graphene and expanded graphite, to supplement the main text conclusion, we focused on their zeta poentials (Fig. S12b). All the GO-based materials exhibited negative zeta potentials > |30 mV|. Such values are considered to be sufficient to provide a stable colloid[39]. In contrast, commercial graphene and expanded graphite, both had well-retained graphene areas, as confirmed by a prominent G peak and a very weak D peak from the Raman analysis (Fig. S12c), but they either floated on or settled in the aqueous solution, failed to form a stable colloidal dispersion in water (inset of Fig. S12b).

It is interesting to note that hydroquinone-reduced GO showed a lower zeta potential (in the absolute value), which was also accompanied by a lower extraction capacity of this rGO (Fig. 2g), as compared to the characteristics observed for ascorbic acid- and hydrazine-reduced GO. These observations indicate that the colloidal stability of rGO influences the accessibility of graphene areas for Au ions.

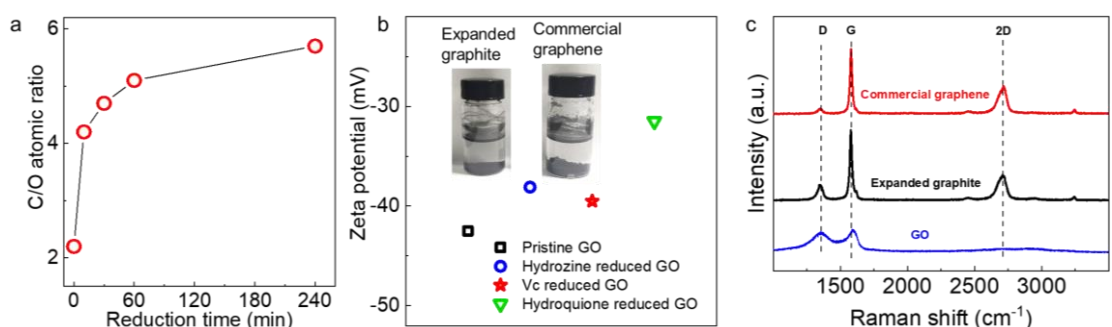

**Fig. S12| Evaluation of different graphene-based adsorbents**. (a) C/O atomic ratio of rGO with

different reduction time using ascorbic acid. (b) Zeta potentials of GO and different rGO suspensions. Insets: photographs show that no stable dispersion of nanosheets of graphene or expanded graphite could be achieved in water. (c) Raman spectra of the GO, expanded graphite and commercial graphene.

#6 Extraction of gold from seawater

As an initial test for the selectivity of rGO, we measured its uptake of metals from an aqueous solution containing 10 ppm of each Au, Cu, Ni and Pt, using salts $KAuCl_4$, $CuCl_2$, $CuSO_4$, $Cu(NO_3)_2$, $NiCl_2$ and $K_2PtCl_4$. As shown in Fig. S13a, rGO allowed recovery of ~99% of Au from the mixture whereas only ~5% Cu, 1.4% Pt and 1% Ni were adsorbed on rGO. Similar % values were also found using 10 ppm solutions of the individual salts rather than their mixture (inset of Fig. S13a), as expected for non-interacting hydrated ions. Not only the metal cationic ions, the above experiments also suggested, there was no noticable influence of the co-existing anionic ions including $Cl^-$, $NO_3^-$, $SO_4^{2-}$, $[PtCl_4]^-$ on the gold extraction performance. This was probably because there was no specific interaction for these anions with graphene or oxidized regions. This behaviour suggests that graphene exhibits preferential affinity to gold as compared to the co-existing ions (Fig. S13b).

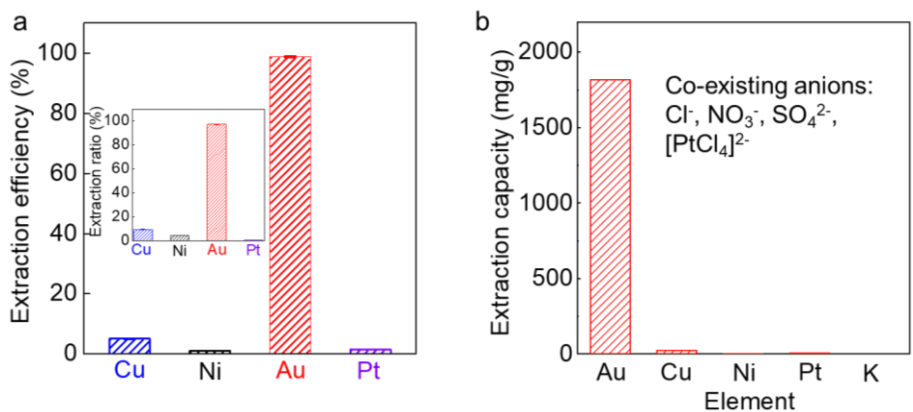

**Fig. S13| Extraction selectivity of rGO.** (a) Extraction efficiency for an equal-part mixture of $[AuCl_4]^-$, $Cu^{2+}$, $Ni^{2+}$ and $[PtCl_4]^{2-}$ ions. Inset: Same for aqueous solutions containing individual metal ions. (b) The influence of the co-existing cations and anions for gold extraction capacity using rGO.

To demonstrate the effectiveness of rGO for gold extraction from solutions containing many different salts, we prepared a simulated seawater solution containing sodium, magnesium, calcium and potassium ions in concentrations typical for oceans. Then 100 ppb of gold ions were added to this solution. First, we carried out the standard extraction protocol from the simulated seawater using pH ≈ 4. The rGO's uptake of gold was found > 99% but with significant presence of Na, Ca and K (Fig. S14). After adding an extra hour at pH ≈ 1 (that is, using protocol 2), the gold uptake increased even further with no noticeable presence of Na and Ca. The remaining K ions (~4% uptake after protocol 2) could be removed by washing away the potassium salts adsorbed on rGO by simply rinsing rGO in water. This shows that the proposed extraction protocols can be adapted to many different situations to achieve a highly selective extraction of Au (Fig. S14).

In further experiments, 0.3 mg of rGO was added to 200 mL of the simulated seawater spiked with $KAuCl_4$ to achieve absolutely minute concentrations of gold ions (10 ppt). After 2 days of extraction, the rGO was collected by centrifugation and drop-casted on a Si wafer for XPS. Similar to the ppt level of gold in pure water, we observed varied gold contents (0.01-0.11 at%) at different areas, which translated into an extraction capacity even higher than the theoretical capacity (calculated based on 100 % extraction efficiency), such high gold content in rGO measured by XPS allowed us to estimate a complete extraction to gold solution at ppt level.

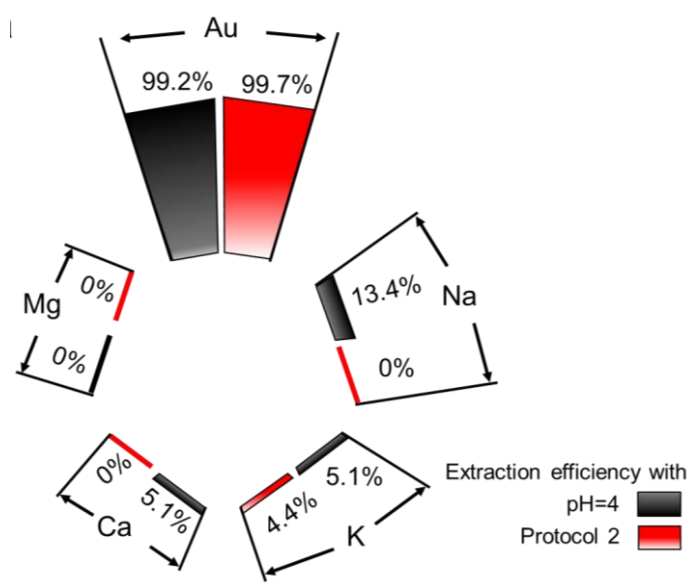

**Fig. S14| Gold extraction from simulated seawater.** Highly selective extraction from the seawater with added 100 ppb of gold. Comparison of efficiencies for different extraction protocols (colour coded).

#7 Gold extraction from e-waste

In our demonstration of e-waste recycling in Fig. 3b, a CPU leachate was diluted to emphasize the superior performance of rGO for extracting even trace amounts of gold and make sure all the gold ions were discharged in the leachate by repeated washing. In real conditions, CPU leachates would contain gold in concentrations from a few to tens of ppm[1,2]. Therefore, we also performed Au extraction from real-world CPU leachates. Specifically, our CPU leachate contained 2.65 ppm Au, 106 ppb tin (Sn), 13 ppb chromium (Cr), 8.5 ppm aluminium (Al), 9.6 ppm Ca, 1 ppb lead (Pb), 54 ppb Ni, 242 ppm Cu, 170 ppb Mg, 68 ppb iron(Fe), 200 ppb zinc (Zn), 70 ppb strontium (Sr), 8 ppb arsenic (As), 620 ppb barium (Ba), and 13 ppb manganese (Mn). Fig. S15 shows that rGO remained very effective with 99% of gold being extracted. Using the pH=4 protocol, rGO also adsorbed approximately 97% Sn, 87% Cr, 62% Al, 39% Pb, 98% As and 2.4% Cu with no discernable adsorption of the other metals present in the leachate. In contrast, if protocol 2 was used, >99% of gold was extracted from the CPU leachate with neither of the 14 coexisting ions being present on rGO (Fig. S15). The ability to achieve such exclusive gold extraction in the presence of many other ions shows an unambiguous potential of the proposed technology for recycling of e-waste.

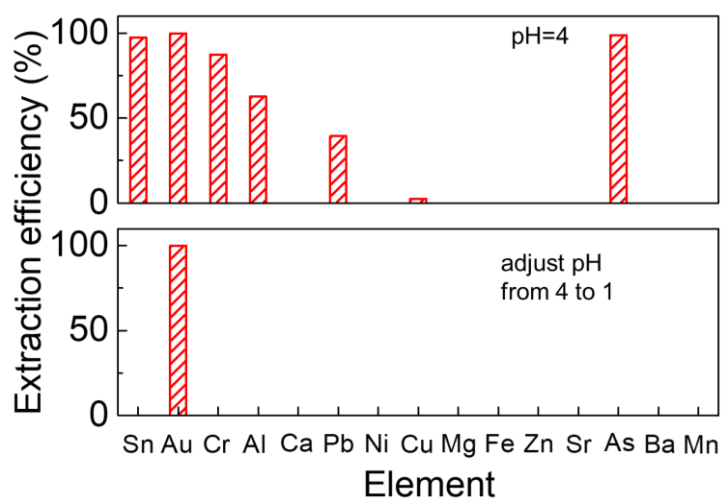

**Fig. S15| rGO performance for gold extraction from real-world CPU leachates.** Top and bottom panels are for different extraction protocols.

#8 Continuous gold extraction using rGO membranes

8.1 Optimal membrane thickness and characterization of extracted gold

To find an optimal thickness of rGO membranes for continuous gold extraction, we measured the solution permeance and extraction efficiencies for membranes having thicknesses of ~ 0.2, 0.8 and 2 μm. 20 ml of a 100 ppm Au solution was filtered through them. For convenience, to measure the relatively high gold concentrations in the feed and filtrate solutions, we used UV-Vis spectroscopy. It showed an absorption peak at ~290 nm, which intensity changed linearly with increasing the gold concentration (bottom inset of Fig. S16a). This peak was then used in real time to determine gold concentrations in the filtrate and to calculate the uptake. Fig. S16a shows that permeance of the membranes decreased with increasing their thickness, as a higher flow resistance is obviously expected for thicker membranes. On the other hand, the extraction efficiency increased with increasing the membrane thickness, which is also expected because thicker membranes get more rGO nanosheets involved in the extraction process. The thickness-dependent trade-off between efficiency and permeance suggests that the continuous extraction process can be adjusted to reach desirable performance by changing the membrane's thickness. Note that, in the described experiments, we did not try to reach highest extraction efficiencies because of rather high Au concentrations and the limited amount of rGO such that the membranes were unable to adsorb all gold present in the tested solution during its single filtration pass.

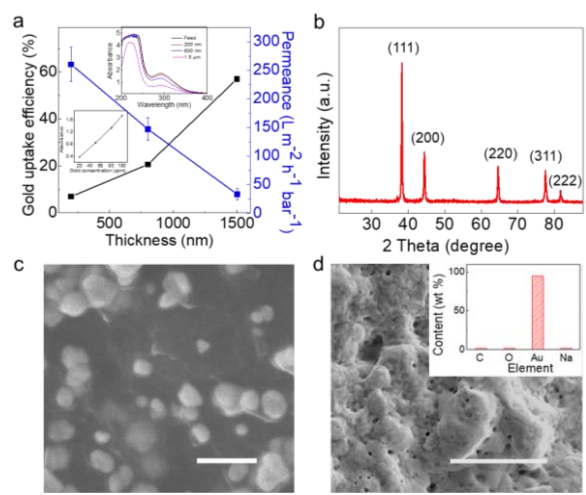

**Fig. S16| Performance of rGO membranes and analysis of the extracted gold.** (**a**) Permeance and gold uptake for membranes with various thickness; 20 mL of a KAuCl$_4$ solution was filtered through. Left inset: intensity of the [AuCl$_4$]$^-$ ion absorption peak at ~290 nm for different Au ion concentrations. The upper inset shows UV-Vis absorption spectra of the feed solution and the filtrate. (**b**) XRD pattern of an rGO membrane (~800 nm thick) after filtration of 6.6 L of a 100 ppb [AuCl$_4$]$^-$ solution. (**c**) SEM micrograph of the membrane's surface after the extraction. Scale bar, 200 nm. (**d**) SEM image of the particulates left after burning off the rGO membrane. Scale bar, 5 μm. Inset: EDS analysis of the particulate.

Next, we filtered a 6.6 L of a dilute [AuCl$_4$]$^-$ solution (100 ppb) through an rGO membrane (3 cm$^2$; 800 nm thick). After the filtration the membrane was exfoliated from its polymer support and studied by SEM. As shown in Fig. S16, many nanoparticles were found on top and inside the rGO membrane. XRD confirmed that they were metallic gold. Burning off the rGO membrane resulted in a deposit containing 95.2 wt% Au, 1.7 wt% Na, 1.67 wt% C, and 1.43 wt% O, as measured by EDS (Fig. S16d). The purity of the resulting gold is calculated to be 23 carats. Based on the TG analysis of pristine rGO (Fig. S4c), the small amounts of detected carbon, oxygen and sodium come not from our extraction process but were probably due to the residual ash of rGO and contamination of the gold surface during its SEM analysis.

### 8.2 Continuous extraction from CPU leachates

In this experiment, we used an rGO membrane to filter a diluted CPU leachate containing 100 ppb of Au. This was designed to check the effectiveness of the proposed continuous extraction rather than to deal with real Au leachates having typically much higher concentrations of Au (see above). Also, we did not carry out this particular experiment in a single filtration step as in Fig. 4. Instead, seven cycles of filtration were used until the 50% permeance was reached (Fig. S17). After each cycle, the membrane was soaked in concentrated HCl at pH =1. Fig. S15 shows the permeance and extraction efficiency over those 7 cycles (140 mL in total was filtrated through). Similar to the case of continuous extraction from solutions containing gold ions only (Fig. 4b), we found that the permeance decreased with each extraction cycle because of the blocking of rGO with metallic gold. After each filtration cycle, we implemented soaking in HCl to desorb coexisting ions from rGO (Fig. S17b). However, this desorption required long time (typically 48 hours) because, unlike rGO colloids, rGO membranes consisted of narrow nanochannels, and it took time for HCl to diffuse and remove adsorbed coexisting ions. Note that the long soaking step could in principle be avoided or shortened by reducing the leachate's pH (soaking and filtration occur simultaneously) or by adding sonication to enhance the diffusion of HCl. Nevertheless, even without further optimization, our results clearly show a possibility of using the proposed rGO technology for continuous gold extraction with exceptional performance.

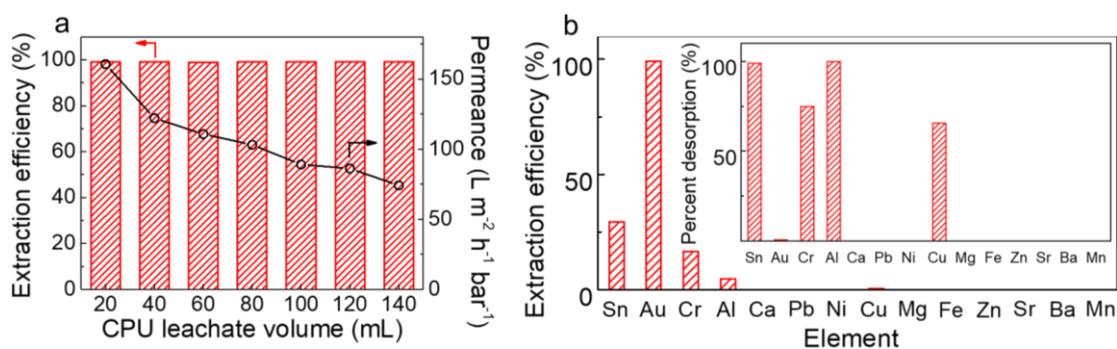

**Fig. S17| Continuous and highly selective extraction of Au from CPU leachates.** (**a**) Changes in extraction efficiency and water permeance through an 800 nm thick rGO membrane during (quasi)continuous filtration. (**b**) Extraction efficiency for different chemical elements present in CPU leachates. Inset: Percentage of coexisting ions removed from the rGO membrane by soaking it in an HCl solution for 2 days.

### 8.3 Recovery of copper from e-waste

After extracting gold from our CPU leachates by continuous filtration, the filtrate contained a significant amount of copper and other coexisting elements. We demonstrate that this copper can also be recovered in a simple process. By adding iron particulates to the leachate remaining after Au extraction, copper inside the solution was reduced (galvanic replacement) and precipitated. The precipitate was collected and analysed by SEM and EDS, which revealed that it contained > 95% of copper (Fig. S18).

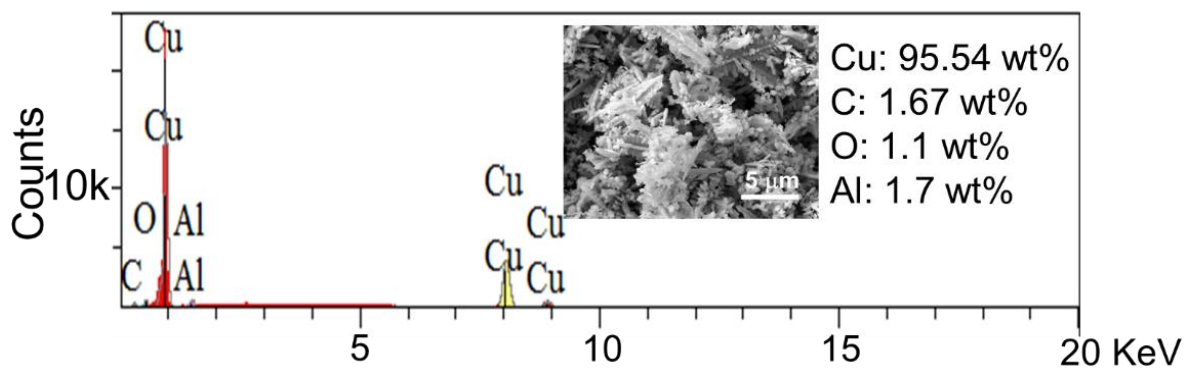

**Fig. S18| EDS analysis of the copper recovered from e-waste.** Inset: SEM micrograph of the recovered copper and its elemental composition.